\documentclass[journal=acscentralscience,manuscript=article]{achemso}

\usepackage{mathtools}
\usepackage{amsmath}
\usepackage{amsthm}
\usepackage{amssymb}
\usepackage{expl3}
\usepackage{anyfontsize}
\usepackage{braket}
\usepackage{relsize}
\usepackage{enumitem}
\usepackage[vlined]{algorithm2e}
\usepackage{algpseudocode}
\usepackage{graphicx}
\usepackage{footnote}
\usepackage{multirow}
\usepackage{tabularx}
\usepackage{booktabs}
\usepackage{makecell}
\usepackage{threeparttable}
\usepackage{color}
\usepackage{natbib,hyperref}
\usepackage{breakurl}

\usepackage{mathtools}
\usepackage{bm}
\usepackage{esvect}
\usepackage{float}
\usepackage{xr}

\makeatletter
\newcommand*{\addFileDependency}[1]{
  \typeout{(#1)}
  \@addtofilelist{#1}
  \IfFileExists{#1}{}{\typeout{No file #1.}}
}
\makeatother

\newcolumntype{C}{>{\centering\arraybackslash}X}

\author{Jie Li*}
\author{Oufan Zhang*}
\author{Yingze Wang}
\author{Kunyang Sun}
\author{Xingyi Guan}
\author{Dorian Bagni}
\author{Mojtaba Haghighatlari}
\affiliation{Pitzer Center for Theoretical Chemistry, Department of Chemistry, University of California, Berkeley, CA, 94720 USA}

\author{Fiona L. Kearns}
\author{Conor Parks}
\author{Rommie E. Amaro}
\affiliation{Department of Chemistry and Biochemistry, University of California, San Diego, La Jolla, CA 92093 USA}

\author{Teresa Head-Gordon}
\affiliation{Pitzer Center for Theoretical Chemistry, Department of Chemistry, University of California, Berkeley, CA, 94720 USA}
\alsoaffiliation{Departments of Bioengineering and Chemical and Biomolecular Engineering, University of California, Berkeley, CA 94720 USA\\
*authors contributed equally}

\email{thg@berkeley.edu}

\title{Mining for Potent Inhibitors through Artificial Intelligence and Physics: A Unified Methodology for Ligand Based and Structure Based Drug Design}
\begin{document}

\begin{abstract}
\noindent
The viability of a new drug molecule is a time and resource intensive task that makes computer-aided assessments a vital approach to rapid drug discovery. Here we develop a machine learning algorithm, iMiner, that generates novel inhibitor molecules for target proteins by combining deep reinforcement learning with real-time 3D molecular docking using AutoDock Vina, thereby simultaneously creating chemical novelty while constraining molecules for shape and molecular compatibility with target active sites. Moreover, through the use of various types of reward functions, we can generate new molecules that are chemically similar to a target ligand, which can be grown from known protein bound fragments, as well as to create molecules that enforce interactions with target residues in the protein active site. The iMiner algorithm is embedded in a composite workflow that filters out Pan-assay interference compounds, Lipinski rule violations, and poor synthetic accessibility, with options for cross-validation against other docking scoring functions and automation of a molecular dynamics simulation to measure pose stability. Because our approach only relies on the structure of the target protein, iMiner can be easily adapted for future development of other inhibitors or small molecule therapeutics of any target protein. 
\end{abstract}


\section{Introduction}
\label{sec:intro}
\noindent
Discovery of new drugs that inhibit target proteins usually follows either a screening-based approach or a rational design approach.\cite{Neves2018,Batool2019} If no explicit information about either the structure of the protein target or a ligand which might bind to the protein is available, the screening of a large database might be the only realistic approach to identify a starting point for drug development. However, any additional information about the structure of either the protein or the ligand can narrow down the search space significantly, and encourages rational ligand-based or structure-based drug design.\cite{Batool2019,Meng2011} The process of rational drug design requires either exquisite domain knowledge and devoted time by an experienced medicinal chemist, or using an automated workflow that relies on virtual screening of protein-ligand  databases combined with physics-based modelling, such as docking simulations, to identify candidate molecules that may potentially bind to the protein target of interest.\cite{Amaro2018}

In order to identify promising small molecule therapeutics, existing high-throughput virtual screening approaches often evaluate comprehensive drug databases such as CHEMBL\cite{Gaulton2012}, PubChem\cite{Kim2021}, and ZINC\cite{Sterling2015}. Even with the recent advent of the Enamine REAL library of 6-11 billion molecules\cite{Grygorenko2020}, it is still dwarfed by estimates for the total number of possible synthesizable small molecules that range from $10^{24}$-$10^{60}$.\cite{polishchuk2013estimation} Unfortunately, due to the size of such established or expanded databases, screening all compounds according to sufficiently sophisticated structure-based methodologies such as flexible ligand docking can be intractable. Even with simpler methods such as pharmacophore modeling or rigid body docking, it can still be time consuming to navigate through the chemically feasible space, with a tendency towards false-positives being ruled in while false-negatives, i.e. potential optimum lead molecules, can be ruled out.\cite{reulecke2008towards,duffy2012early}

With the rise of modern machine learning, a more clever exploration of the chemical space for drug-like molecules becomes possible, thanks to the development of generative models that can generate molecules represented as either strings or graphs.\cite{sanchez2018inverse,kusner2017grammar,dai2018syntax, subramanian2020inverse,olivecrona2017molecular,popova2018deep,gottipati2020learning,zhavoronkov2019deep,zhavoronkov2020potential,bung2021novo,born2021data} More interestingly, the distribution can be skewed towards molecules with specific properties such as drug likeness using techniques such as variational autoencoders (VAE)\cite{kusner2017grammar,dai2018syntax}, transfer learning \cite{subramanian2020inverse} and reinforcement learning (RL)\cite{olivecrona2017molecular,popova2018deep,gottipati2020learning,zhavoronkov2019deep,zhavoronkov2020potential,bung2021novo,born2021data}. However, early deep learning methods rely on 1-dimensional (1D) sequence or 2-dimensional (2D) chemical representations of the drug and protein, and do not take full advantage of 3-dimensional (3D) structural information of the putative drug, thereby constraining the ability to \textit{generate} drugs with shape and molecular compatibility with the target active site. When SMILES-based generative models are combined with additional information about protein structure, ligand structure or protein-ligand interactions, the quality of generated molecules improved in terms of predicted binding affinity or binding mode similarity with a reference binder.\cite{zhang2022novo,li2023ls} Recent work has explored chemical space in the vicinity of some starting molecular scaffold and running docking simulations on these derived molecules\cite{jeon2020autonomous}, but the chemical space that can be explored by such a method is still rather limited. Generation of atomistic structures of new compounds conditioned on the 3D structure of a provided pocket or constrained on interaction fingerprints has also been proposed  with sequential growth of atoms\cite{pocket2mol}, hierarchical buildup of 3D coordinates \cite{zhang2023resgen} or generating all-atom coordinates at once using diffusion models\cite{schneuing2022structure}. However, these existing methods have either been tailored towards \textit{de novo} generation or a local optimization, but lack the flexibility to extend from \textit{de novo} design to structure or ligand based rational design. These additional capabilities are essential for continuously improving the binding potency between ligands and proteins after a hit molecule is already identified. 

In this work we propose a novel composite workflow, dubbed "iMiner", that mines chemical space for new tight binding inhibitors by combining deep RL with real-time flexible ligand docking against a protein binding site. We represent putative inhibitors as Self-Referencing Embedded Strings (SELFIES)\cite{Krenn2020} that are generated from an Average Stochastic Gradient Descent Weigh-Dropped Long Short Term Memory (AWD-LSTM)\cite{Merity2018} recurrent neural network (RNN), allowing wide coverage of chemical space. We illustrate the RL training procedure of iMiner that uses on-the-fly AutoDock Vina\cite{Trott2010} in a predefined binding pocket of the 3D structure of the protein to generate small molecule inhibitors, in which the AutoDock-vina score is used to adjust the RNN so that the distribution of generated inhibitor molecules are shifted towards those that more strongly interact with the protein. The algorithm also allows for users to adapt iMiner to satisfy additional restraints such as encouraging 2D ligand similarity to vary the chemistry of a known inhibitor, to grow molecules from a 3D fragment positioned in the pocket, as well as enforcing ligand interactions with particular protein interaction sites. Finally, iMiner also offers the option to provide a series of post-filters to downselect the generated molecules so that they obey Lipinski rules\cite{Lipinski2000}, increase the likelihood of synthetic accessibility (SA) and drug likeliness, avoidance of non-specifc binders (Pan-assay interference compounds or PAINS)\cite{dahlin2015pains} and find consensus molecules with alternative docking software and scoring functions.

To validate the effectiveness of the iMiner approach, we have designed inhibitors for the popular SARS-COV-2 Main protease (Mpro) as an example. We chose this system because of the ready availability of experimental 3D structures\cite{jin2020structure} and the fact that multiple effective ligands have already been proposed to inhibit Mpro\cite{zhang2021potent}, which provides ample information for different structure-based and ligand-based designing strategies.\cite{mpro_review} We illustrate the four features of the iMiner approach and evaluate our model's capability to generate new inhibitors, perform structural modifications while keeping the pharmacophore intact, growing ligands from a fragment starting point, or retaining key interactions with a specified protein residue. Although illustrated on the SARS-COV-2 Mpro target, we would like to emphasize that the iMiner workflow can be readily adapted to generate small molecules for other protein targets, since it only requires a 3D structure of the target protein with a pre-defined binding site. Thus, we believe our ML algorithm will be of great interest to the drug design community to rapidly screen novel regions of chemical space in real-time for other anti-virals or small molecule therapeutics in the future. 

\section{The iMiner Algorithm and Workflow}
\noindent
Figure \ref{fig:workflow} provides an illustration of the overall structure of the iMiner algorithm which highlights the
two major machine learning components, the generative and evaluative models and their interplay, for generating new inhibitor molecules. In addition we embed the RL-physics model into a composite workflow for further analysis and down-selection, also displayed in Figure 1. Here we describe the iMiner algorithmic components and workflow in more detail. 

\subsection{The Generative Module}
\begin{figure}
\begin{center}
\vspace{-15mm}
\includegraphics[width=0.9\textwidth]{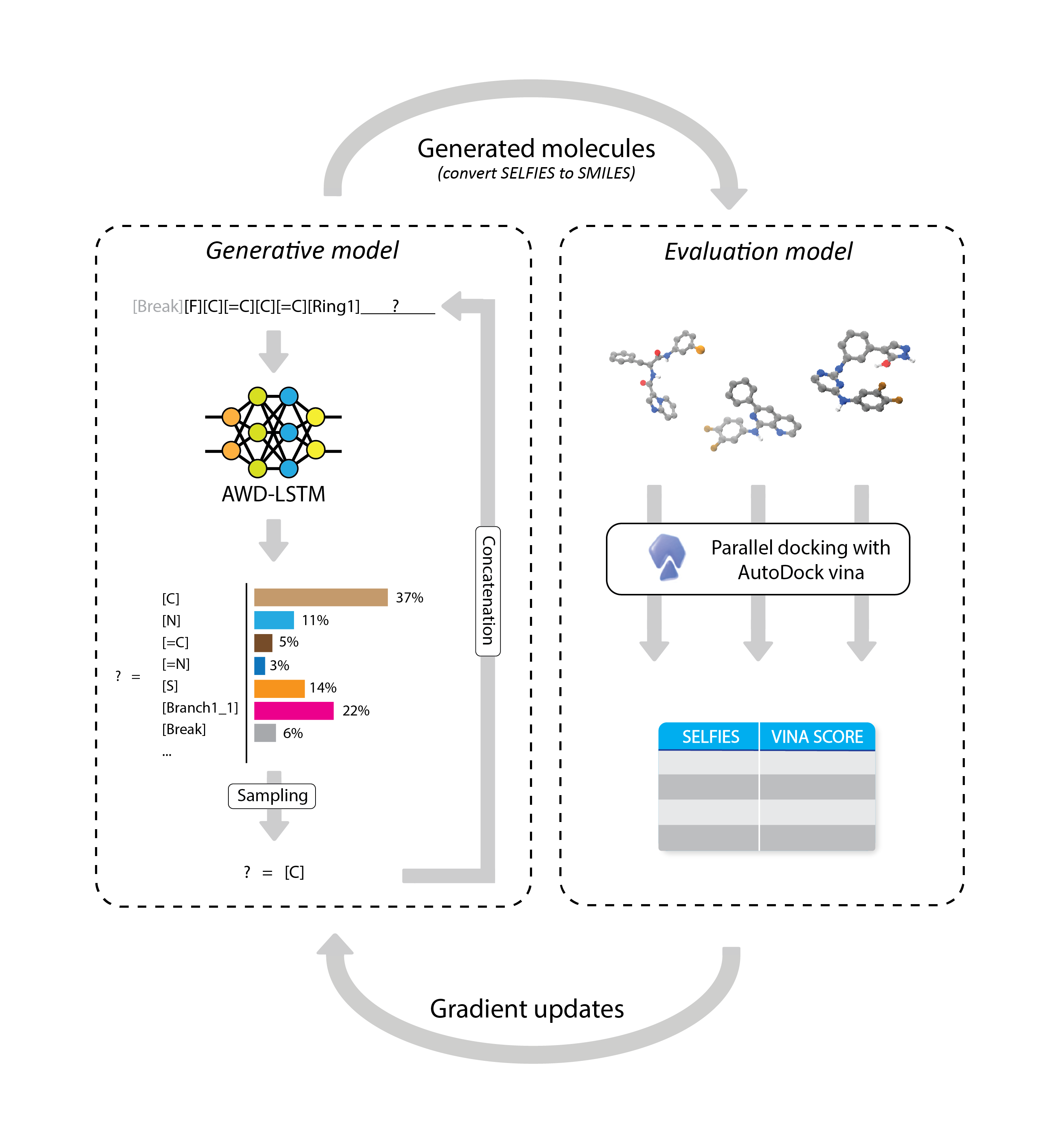}
\includegraphics[width=0.8\textwidth]{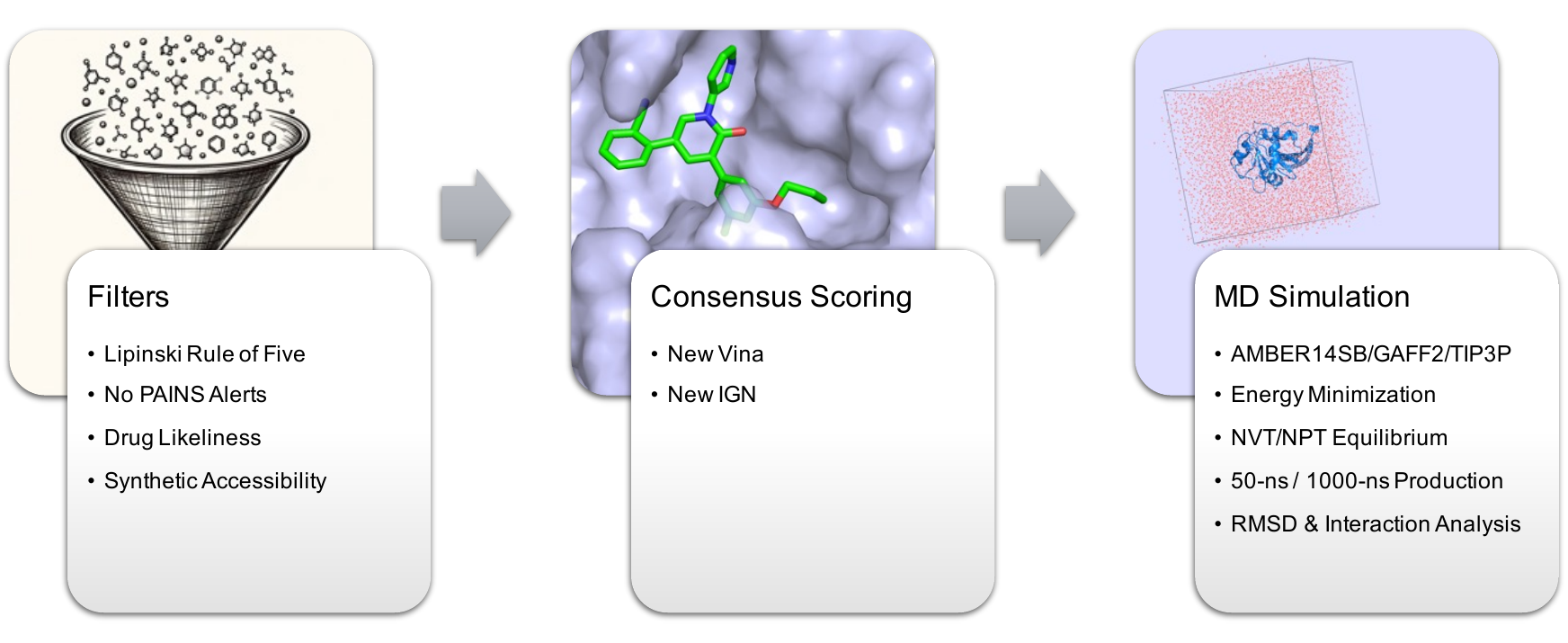}
\end{center}
\vspace{-4mm}
\caption{\textit{Illustration of the overall structure of the iMiner algorithm and workflow}. (a) The generative model uses SELFIES representations for molecules and a recurrent neural network to ``mine" for new molecules that are presented to the evaluative model for 3D docking using AutoDock Vina. Vina scores and other loss/reward functions are used to drive gradient updates of the neural network. (b) The iMiner workflow provides automated post-selection filters, consensus scoring and MD simulations for overall ranking.}
\label{fig:workflow}
\end{figure}

\noindent
Conceptually, generating molecules using a string representation is similar to how text is generated in a natural language processing task. Our method starts with a specific {\fontfamily{qcr}\selectfont [Break]} token, and for each molecule, we utilized an RNN that takes in the last token in the string, together with the hidden state from last step to predict a distribution of tokens following the current string. In this work a specific variant of the RNN, known as the AWD-LSTM, was used due to its exceptional performance in similar generative tasks (Figure \ref{fig:workflow}a).\cite{Merity2018} 

\textbf{SELFIES representation of inhibitor molecules}. An arbitrary molecule can be represented as a topological graph using two main approaches: adjacency matrix methods and string based methods. The former uses an N by N matrix to encode a molecule, where N is the number of atoms in the molecule, and the values of the matrix are typically bond orders between atoms. An adjacency matrix is not ideal for generative tasks, because the 

\noindent
size of the molecule that can be generated should not be fixed, and the learning of chemical knowledge by a ML model through adjacency matrix can be difficult. Instead, string based methods are more suited for molecular generation tasks, and SMILES strings have been the standard for molecular representation due to its conciseness and readability. However, SMILES strings have relatively complex syntax, require matching of open and close brackets for branching, and ring modeling/modification is not trivial. Therefore, generating novel, chemically correct compounds through use of SMILES strings can be challenging.

The SELFIES molecular representation\cite{Krenn2020} is specifically designed to ensure that all generated strings correspond to valid molecules. By utilizing {\fontfamily{qcr}\selectfont [Branch]} and {\fontfamily{qcr}\selectfont [Ring]} tokens with predefined branch lengths and ring sizes, as well as generating symbols using derivation rules, the SELFIES representation guarantees that valence bond constraints are met, and any combinations of tokens from its vocabulary corresponds to a valid molecule. Therefore, we have used SELFIES in our generative model to encode molecules since it does not need to learn chemical syntax rules, and can allocate more of its learning capacity towards generating valid molecules with properties of interest.

\textbf{Pre-training the inhibitor molecule generation}. The network was pre-trained using supervised-learning (SL) of all molecules from the ChEMBL database to learn the conditional probability distributions of tokens that correspond to drug-like molecules. When our trained generative model is used for generating new molecules, a new token is sampled according to the predicted probabilities, and this new token is concatenated to the input string to sample the next token, until the {\fontfamily{qcr}\selectfont [Break]} token is sampled, in which case a complete molecule has been generated.

We then validated our pre-trained distributions using 13 drug-likeliness properties between our generated molecules and randomly sampled molecules from ChEMBL database that we used for training. The molecular properties considered are well-recognized chemical features related to the drug-likeliness of a molecule which can be obtained through 2D topological connectivity of the molecule: fraction of $sp^3$ hybridized carbons, number of heavy atoms, fraction of non-carbon atoms in all heavy atoms, number of hydrogen bond donors and acceptors, number of rotatable bonds, number of aliphatic and aromatic rings, molecular weight, approximate log partition coefficient between octanol and water (alogP)\cite{wildman1999prediction}, polarizable surface area (PSA), the number of structural alerts\cite{brenk2008lessons}, and the size of the largest ring in the molecule. Despite the fact that during pre-training only token distributions were used as training targets, all distributions collected from our generated molecules closely follow the distributions from the ChEMBL database (Figure S1). This result suggests our pre-trained model has learned key concepts of ``drug-likeness" and provides a good starting point for the RL procedure.

\subsection{The Evaluation Module}
After our generative model was pre-trained, we employed an RL workflow to bias the distribution of generated molecules towards specific properties of interest. RL training allows the model to interact with an environment by performing actions according to a policy model, and uses the feedback from the environment to provide training signals to improve the model. In this work, the pre-trained generative model is taken as the policy, and in each iteration 2000 molecules were generated and sent to the evaluation module (Figure \ref{fig:workflow}a). 

\textbf{Physics-based Docking}. The central component of our evaluation model is docking with AutoDock Vina (Vina) in parallel with the RL. Within our evaluation model, the Vina score calculator take a SMILES string representing the ligand, and the 3D structure of the protein target, together with a predefined docking region as input. AutoDock Vina then explores variations of the dihedral angle degrees of freedom and identifies the optimal conformation of the input inhibitor for placement in the designated protein binding site. Finally, AutoDock returns the Vina score as an approximation of the binding energy between the ligand and the protein. Multiple instances of the Vina score calculator tasks were established through GPU parallelization to allow high-throughput evaluation of the generated molecules.\cite{vina_gpu} Vina scores were then cycled back to the generative model to improve molecule generation through proximal policy optimization (PPO)\cite{schulman2017proximal}, as will be discussed in next section. We emphasize that by using a physics-based docking model which utilizes full 3D structure of our target protein and generated molecules as the critic, the training of the policy model is less likely to be contaminated due to exploiting failure modes of a neural-network based critic, an issue called wireheading\cite{everitt2016avoiding}. Instead, we benefit from a more reliable training signal and reduce the false positive and false negative rates of the generated molecules.

Vina scores alone are not sufficient to reliably train a molecule generator, as shown in the Supporting Information (Figure S2) because it will not always satisfy requirements for drug-likeness. To ensure that our generated molecules still bear drug-like properties, we incorporated an additional metric into the reward, $S_{DL}$, which is a weighted average of the log likelihood for the 13 different drug-like properties used in pre-training assessment. Our custom drug-likeliness score is an extension of the widely used quantitative estimate of drug-likelihood (QED) value\cite{bickerton2012quantifying} tailored to our deep learning generative model. Formally, the drug-likeliness score $S_{DL}$ is defined as:
\begin{align}
     S_{DL}(\bm X)=\sum_{i}{\sigma_i \log{p_i(\mathrm{prop}_i(\bm X))}}
\end{align}
where $\mathrm{prop}_i(\bm X)$ calculates the $i$th property of a molecule $\bm X$ and $p_i$ is defined by the probability distribution of property $i$ by all molecules in the ChEMBL database. The parameter $\sigma_i$ is defined as:
\begin{align}
    \sigma_i=S_i^{-1}/\sum_j{S_j^{-1}}
\end{align}
where $S_i$ is the entropy of the distribution of property $i$,
\begin{align}
    S_i=-\sum_x{p_i(x) \log{p_i(x)}}
\end{align}
such that a narrower distribution from the ChEMBL database contribute more to the drug likeliness score, and defines the weights for each property as proportional to the inverse of the entropy. Introducing this additional reward ensures our model also accounts for similarity of generated molecules to the drug-likeliness present in the ChEMBL database, and ensures that our generated molecules are more likely to be optimal leads for further drug design endeavors.

\textbf{Reinforcement learning with different reward types}. Our pretrained policy model defines a probability distribution for an arbitrary sequence of tokens from the SELFIES vocabulary, since the generation of the sequence is a Markovian decision process (MDP), and can be written as:
\begin{align}
    p_\Theta(s_T)=p_\Theta(s_1|s_0) p_\Theta(s_2|s_1)... p_\Theta(s_T|s_{T-1})
\end{align}
where $s_0$ corresponds to a starting state with {\fontfamily{qcr}\selectfont [Break]} as the only token in the string, $s_t$ corresponds to an intermediate state with a finite length string of SELFIES tokens not ended with the {\fontfamily{qcr}\selectfont [Break]} token, and $s_T$ corresponds to the terminal state, with the last token being {\fontfamily{qcr}\selectfont [Break]}, or the length of the string exceeding a predefined threshold. $p(s_t|s_{t-1})$ is the transition probability at time step $t$, which is the probability distribution of the next token from the generative RNN with network parameters $\Theta$. For each terminal state not exceeding the length limit, a corresponding molecule can be decoded, and the Vina score $S_{vina}$ and drug-likeliness score $S_{DL}$, can be calculated and further optimized. The total reward for a terminal state with a decoded molecule $\bm X$ is then defined as:
\begin{align}
    r(s_T)=\sum_i{\lambda_i R_i(\bm X)}
\end{align}
where $i$ denotes the reward function types, and $\lambda$ parameters control the balance between the different scores.
For the unconditional \textit{de novo} generation case, where only the drug-likeliness score needs to be maximized and Vina score needs to be minimized, the reward becomes 
\begin{align}
    r(s_T)=\lambda_{DL}\max(S_{DL}(\bm X),0)-\min(S_{vina}( \bm X),0)
\end{align}
Negative $S_{DL}$ is upward clipped to 0 and positive $S_{vina}$ is downward clipped to 0 to ensure the reward is non-negative. The expected reward under the MDP is then 
\begin{align}
    J(\Theta)=\mathbb{E}_{s_T\sim p_{\Theta}(s_T)}[r(s_T)]
    \label{formula:rl_target}
\end{align}

To encourage the model to generate molecules based on a certain scaffold, we introduced a fragment similarity score, defined by the maximum Dice similarity of the Morgan fingerprints between fragments of the query molecule decomposed using BRICS\cite{BRICS} and the target molecule or structural components, as well as a pharmacophore similarity score with the Pharm2D\cite{Gobbi1998} module in RDkit\cite{Landrum2016RDKit2016_09_4}. In addition to molecular similarity, we also implemented a positional penalty of the docking score to impose 3D structural similarity if needed. The algorithm identifies the most similar fragment component in a sampled molecule, calculates the shape alignment of the component to the target fragment and clips the docking score to 0 if the alignment difference exceeds a given threshold. 

A further augmented feature of iMiner is to generate  molecules that interact with specific protein residues in the pocket. We developed an interaction score as the weighted total of interactions of specific types between predicted binding pose of a sampled molecule and the specific residues using Protein-Ligand Interaction Profiler (PLIP)\cite{Salentin2015}, where the weights are defined by users to prioritize certain interaction types and/or protein residues as desired. We also differentiated hydrogen bonding with angle \textless 135$^\circ$ and the hydrogen-donor bond length \textgreater 2.8 \AA  as weak interactions and down-scaled the corresponding interaction scores by 0.5 to encourage the generated molecules to form stronger bonds. 

While we present results below using the different reward scores separately for clarity, all the above customized scores can be used jointly along with drug-likeliness score and Vina score to meet different design criteria. Further details of the RL training procedure and hyperparameters $\lambda$ are given in the Methods section and Supplementary Information.

\subsection{The iMiner Workflow} 
The iMiner algorithm is embedded in a composite workflow that automates the post-analysis of generated molecules to help with overall ranking (Figure \ref{fig:workflow}b) After completion of molecule generation and optimization, it is desirable to filter out PAINS molecules as well as molecules with Lipinski rule violations\cite{lipinski1997experimental}, and to select for molecules with good synthetic accessibility (SA) scores. We therefore developed and applied a set of filters requiring no PAINS, no Lipinski rule violations, and SA scores using RDKit package\cite{Landrum2016RDKit2016_09_4}, as well as a drug-likeliness score filter of \textgreater 2.8 to exclude any structure that is not sufficiently drug-like. For the task to generate molecules with desired protein-ligand interactions, we applied an additional filter to extract molecule with confirmed interactions with the specified protein residue(s). While we use these specific values for the results described below, users have the option to change these values if desired.

Checking for consensus with alternate scoring functions is often considered good practice as any individual scoring function may have limited accuracy or be parameterized for different test cases.\cite{houston2013consensus} Thus the workflow collects all molecules from RL training iterations with a user-defined Vina score cutoff, and evaluates them with two alternative scoring functions. For cross-validation we utilized a new AutoDock Vina score (newVina) and new InteractionGraphNet (newIGN)\cite{IGN} that were recently retrained using LP-PDBBind\cite{LP-PDBBind}, a cleaned PDBBind data that implements control on the data similarity between the train, validation and test dataset. The final step was to rescore the above resulting molecular poses with the newVina and newIGN scores, and extract molecules that are in a certain consensus top range.

The final stage of the iMiner workflow is to run molecular dynamics (MD) simulation for the protein-ligand complex. It has been widely recognized that molecular docking has some difficulties in terms of distinguishing non-binders from binders as a consequence of lacking ensemble information and inability to capture induced-fit effects. Physics-based MD simulation have been proven to resolve such issues to some extent by effectively incorporating flexibility of both the receptor and ligand while also including molecular aqueous solvent. Therefore, we incorporate MD simulation into iMiner as a tool to validate the protein-ligand binding stability. 

The automated MD procedure is described as follows: (1) the protein-ligand pose with the best Vina score is used as the initial structure and the simulation system is prepared by adding water solvent molecules and when required neutralizing counter ions ($\mathrm{Na^+}$ or $\mathrm{Cl^-}$); (2)  AMBER14SB/GAFF2/TIP3P are adopted to parameterize the protein, ligand and water respectively; (3) the system is first energy minimized followed by a 1 ns equilibration simulation in the NVT ensemble under 1 bar, 298.15 K, followed by a 1-ns simulation under the NPT ensemble; (4) a 50-ns production run is performed in the NPT ensemble and a total of 250 structures are collected for analysis in step 5; (5) for each structure the ligand's RMSD is calculated with respect to the starting pose, and the probability of interactions between the ligand and protein are analyzed with PLIP for each structure.\cite{Salentin2015}; (6) For stable ligand-proteins from step 5 we perform a long production run in the NPT ensemble for 1 $\mu$s.

\section*{Results}

\subsection{Unconditional \textit{de novo} generation}
\begin{figure*}
\centering
\includegraphics[width=0.9\textwidth]{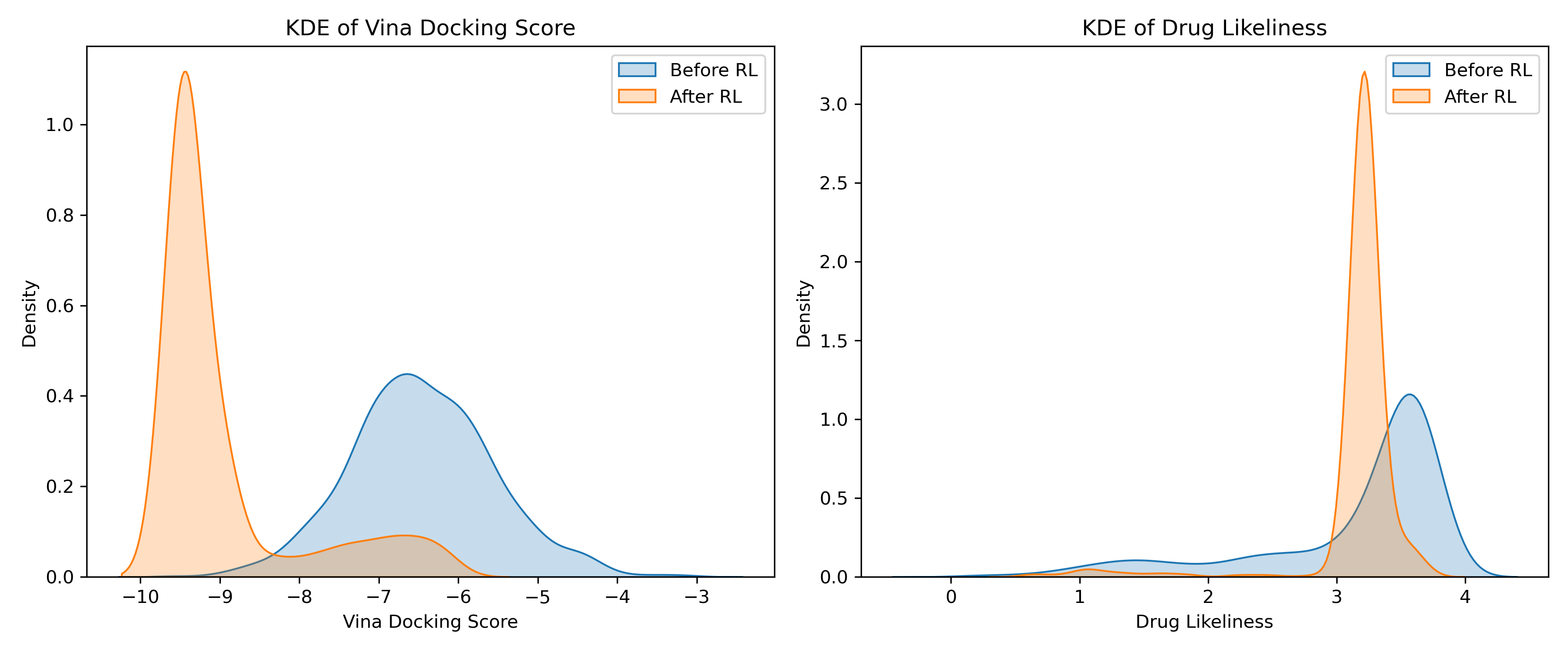}
\caption{\textit{Kernel density estimate plots of vina docking score and drug likeliness for de novo generated molecules before and after RL training.} For the comparison, 1000 molecules were generated using the models before and after RL training to create the plot.}
\label{fig:task-1-distribution}
\end{figure*}

We use the SARS-CoV-2 Mpro as an example to demonstrate iMiner's ability to generate a diverse set of potential drug-like small-molecule inhibitors without any information beyond the active site pocket. To achieve this goal of de novo generation, two objectives were optimized during the RL training: (1) AutoDock Vina docking score to ensure the proposed molecules have good binding efficacy, and (2) drug-likeliness to constrain our chemical search space only to molecules that are considered as drug-like. In Figure \ref{fig:task-1-distribution}, we compare the distribution of Vina score and drug likeliness score before and after RL training. The clear shift towards more negative Vina scores while maintaining a similar drug-likeliness distribution shows that iMiner can bias the generative model towards more potent and drug-like binders. For de novo generation of molecules against a given target, another important aspect is whether it is generating unique molecules consistent with the 3D target active site, at least within the realm of validity of the AutoDock Vina model. 

To quantify the generated molecule distributions before and after the RL optimization, in Table \ref{tab:guacamol} we evaluate the GuacaMol benchmarks\cite{brown2019guacamol}, which probe 5 different aspects of the distribution of generated molecules with respect to the training dataset. Model "validity" reports the proportion of molecules that are syntactically correct. Because we generated molecules via SELFIES representations, we achieved close to 100\% validity for all generated molecules in both the pre-trained and after the RL optimization. Invalid molecules were either empty strings, or molecules for which the SELFIES package failed to convert into a SMILES string, and therefore were discarded before the next workflow steps. Model ``uniqueness" reports how many generated molecules are duplicates versus those which are genuinely distinct. Our pretrained and RL models illustrate high uniqueness, indicating the model is able to generate a wide variety of non-redundant molecules. Model "novelty" is defined as the proportion of generated molecules that do not exist in the training dataset. Both our pre-trained and RL model's high novelty indicates that it is not memorizing molecules from the training dataset, but is indeed generating molecules that it has not seen before. By these 3 metrics the pre-trained and RL molecules satisfy the criteria of an unconditional de novo generation of unique molecules.

\begin{table}[h]
    \centering
    \caption{\textit{GuacaMol benchmarks for the pretrained generative model and after RL training.} The model benchmarks include valid chemical molecules, uniqueness and novelty with respect to the training set, and distribution similarity evaluated using KL diverhence and Frechet ChemNet distance. }
    \begin{tabular}{lcc}
        \hline\hline

            \textbf{Benchmark}& Pretrained model & After RL     \\ \hline
        \textbf{Validity} & 0.998  & 0.998 \\ 
        \textbf{Uniqueness} & 0.999 & 0.983  \\
        \textbf{Novelty} & 0.867& 0.999 \\
        \textbf{KL divergence} & 0.985 & 0.791 \\
        \textbf{Frechet ChemNet Distance} & 0.870 & 0.007\\
                                          
        \hline
    \end{tabular}

\label{tab:guacamol}
\end{table}

But the RL learning should also generate unique molecules that are specific to the 3D interaction space of the binding pocket. The Kullback–Leibler (KL) divergence measures differences in probability distributions of various physicochemical descriptors for the training set and the pre-trained and RL model generated molecules. As defined by GuacaMol, a high KL divergence benchmark suggests that the generated molecules have similar physicochemical properties to that of training dataset. This is true for the pre-trained models by design, but what is clear after the RL optimization is that the influence of the 3D shape requirements of the protein pocket and emphasis on drug likeness results in significant deviation from the original training set. This is also reflected in the Frechet ChemNet Distance (FCD), which utilizes a hidden representation of molecules in a previously trained NN to predict biological activities, and thus captures both chemical and biological similarities simultaneously for two sets of molecules.\cite{preuer2018frechet} Molecules generated by our pre-trained model have high FCD values, indicating that our molecules are expected to have similar biological activities as molecules from the ChEMBL training dataset, but the strong deviation after RL training are molecules more specific to the SARS-Cov-2 Mpro target.

\begin{figure}
\centering
\includegraphics[width=0.9\textwidth]{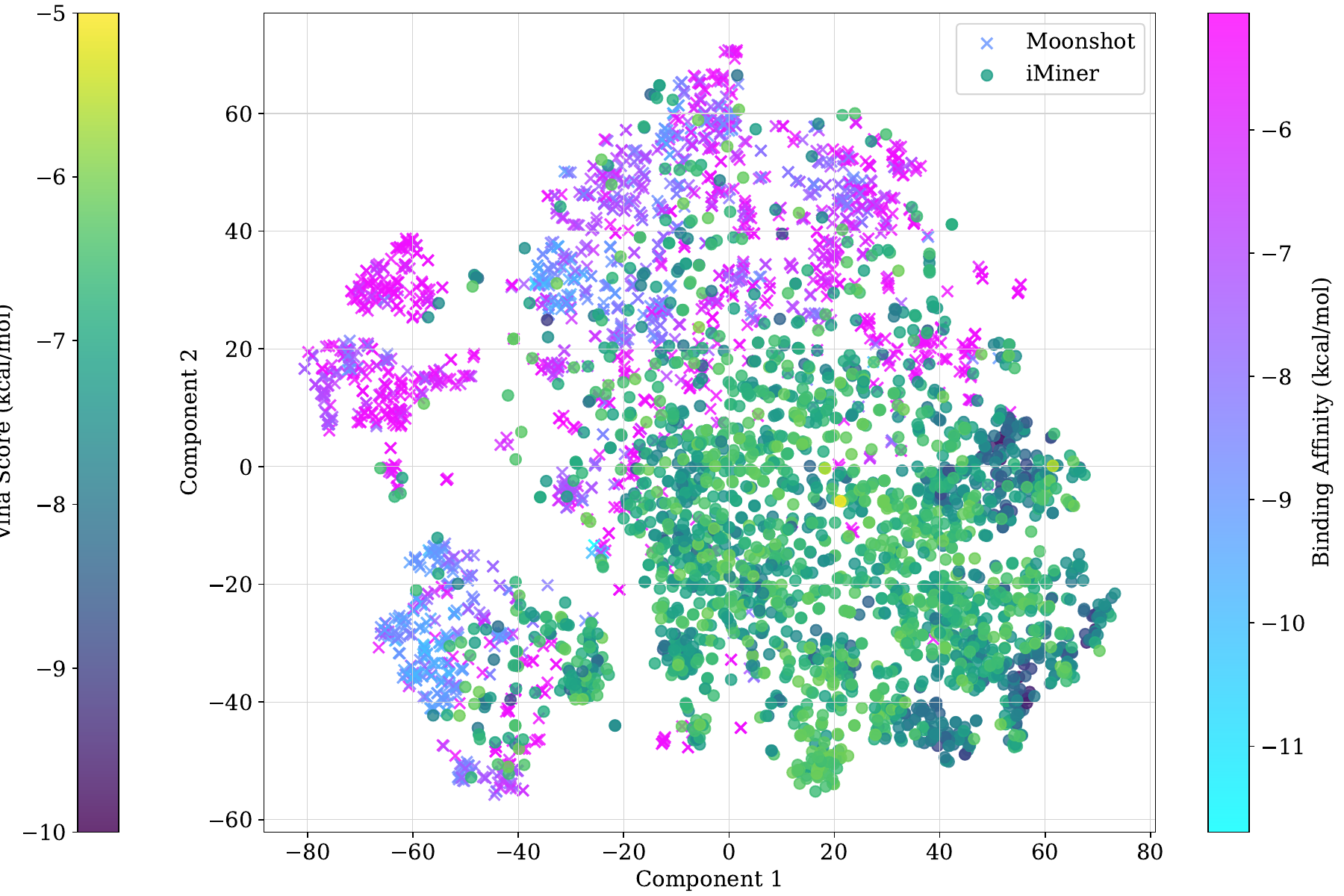}
\caption{\textit{Visualization of molecules from the COVID moonshot project (crosses), molecules generated and post-filtered by the iMiner algorithm (dots).} Molecules are encoded into a 512-dim vector with ChemNet\cite{preuer2018frechet} approach and reduced to 2-dim with t-distributed stochastic neighbor embedding (t-SNE). Moonshot molecules are color-coded by its binding affinity (converted from IC50 values) and iMiner-generated molecules are color-coded with the unmodified AutoDock Vina score.}
\label{fig:task-1}
\end{figure}

Another metric for defining ligand uniqueness specific to SARS-Cov-2 Mpro is to compare the iMiner results to that of the COVID Moonshot project\cite{boby2023open}. After post-filtering with the aforementioned procedure (in this task we used a Vina cutoff percentage of 20), 2347 molecules were collected from all RL iterations, and each of the molecules is represented by a 512-dimensional vector calculated with the ChemNet algorithm\cite{preuer2018frechet} and then reduced to a 2-dimensional vector with t-distributed stochastic neighbor embedding (t-SNE) approach for better visualization. Within these principal components, Figure \ref{fig:task-1} compares the coverage of chemical space of the iMiner molecules with inhibitors identified in the COVID Moonshot project\cite{boby2023open}. 

While there is some overlap between the iMiner and Moonshot molecules both in terms of similarity and binding affinity, it is most notable that the iMiner-generated molecules covers a quite different chemical space than molecules published in the Moonshot project. The clustering of the COVID moonshot molecules stems from the fact that many are generated through an inspirational approach, i.e., later molecules are borrowing design  ideas and sub-structures from molecules submitted earlier. By comparison, the iMiner molecules are dispersed throughout chemical space and provides a wide variety of structures as candidates for lead optimization. As drug leads built on single or closely related scaffolds might be ruled out entirely during later drug development stages, a wider coverage of the chemical space offers the possibility of new Mpro inhibitors.

\begin{figure}
\centering
\includegraphics[width=0.92\textwidth]{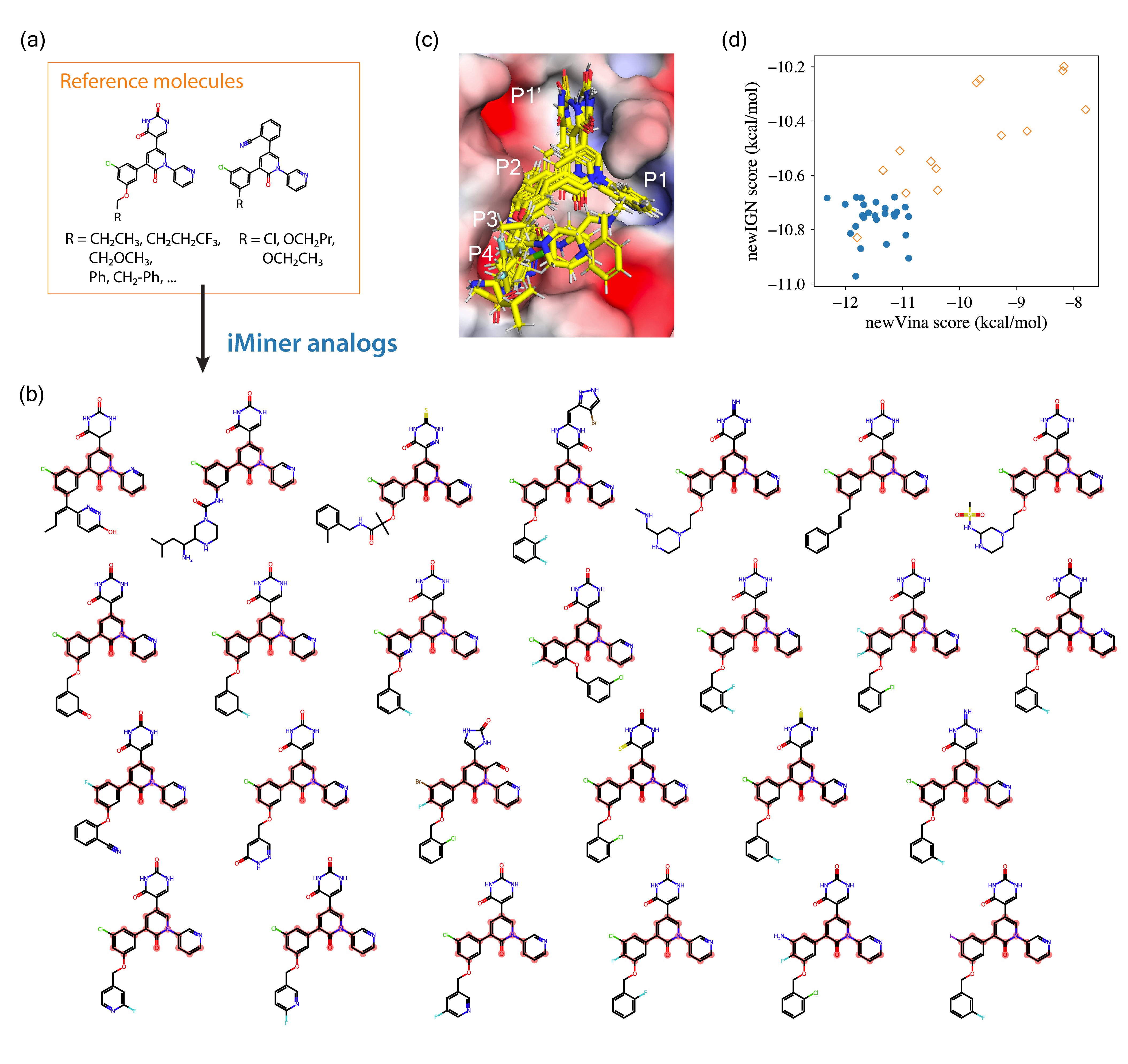}
\caption{\textit{iMiner generated structural analogs from a set of reference molecules.} a). The perampanel derived \cite{zhang2021potent} reference molecules, containing a central pyridinone ring connecting a pyridinyl group and a meta-substituted phenyl ring. b). 2D visualization of the iMiner generated analogs with the shared motif highlighted in pink. c). Surface rendering of the iMiner generatede analogs in the Mpro canonical binding pockets. d). The newVina and newIGN scores of the Autodock Vina docked poses of the analogs generated by iMiner (blue) and the reference molecules (orange).}
\label{fig:analog}
\end{figure}

\subsection{Generation of structural analogs}
Structural analogs are designed to have a similar structure to a reference molecule which is (often) a known binder, but made to differ in certain functional groups to optimize the structure-activity relationship (SAR). Developing structural analogs are valuable in drug discovery because they can display a spectrum of biological activities distinct from the parent molecule, but it remains a combinatorial search challenge such that automated computational approaches can be of great value. We therefore have adapted iMiner's reinforcement learning coupled with docking algorithm to also include a  structural similarity score to efficiently explore the variations in chemical space around a given target compound. We have considered two types of structural similarity: fingerprints that encodes the topology of a molecule, as well as pharmacophore models that capture the arrangement of chemical features important for biological activity. In this work, we exemplify this approach by generating analogs of the SARS-CoV-2 Mpro inhibitors reported by Zhang \textit{et al.}\cite{zhang2021potent}. 

Figure \ref{fig:analog}a displays the reference molecules which are comprised of a core cloverleaf motif around a central pyridinone ring, a pyridinyl group and a meta-substituted phenyl ring directed towards His41, features considered to be well-optimized in ref\cite{zhang2021potent}. As seen in Figure \ref{fig:analog}b, the iMiner molecules generated to maintain similarity to the core motif have also varied the functional groups extending into the P3-P4 region of the Mpro canonical binding pockets as seen in Figure (\ref{fig:analog}c. In addition, Figure \ref{fig:analog}d show that both newVina and newIGN scores of the docked poses of the generated molecules determined by Autodock Vina aggregate in the more negative energy (higher affinity) range compared to those of the reference perampanel derivatives, since the iMiner model also optimizes binding affinities while exploring a local chemical space around the reference molecules. 

\begin{figure*}[tbp]
\centering
\includegraphics[width=0.92\textwidth]{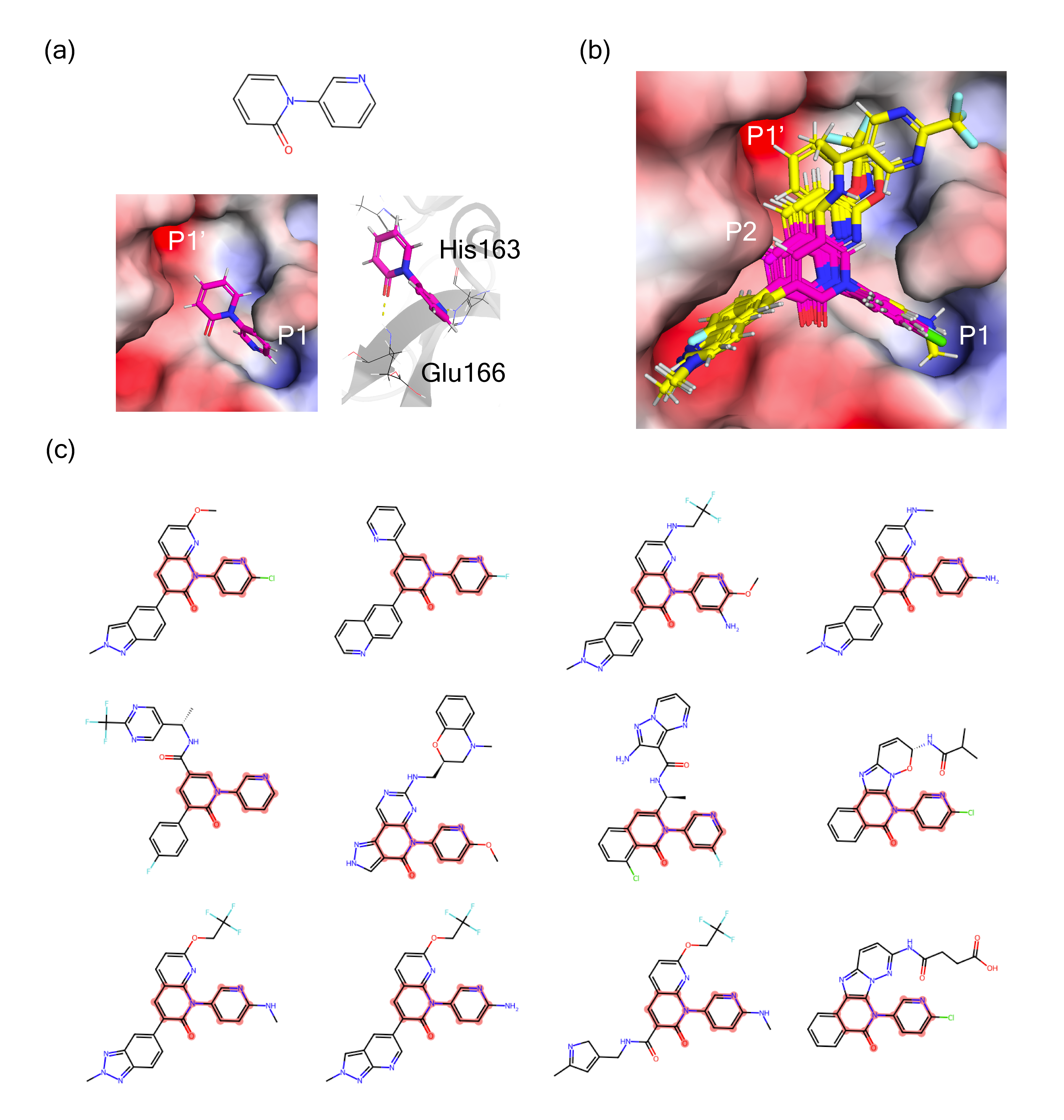}
\caption{\textit{Structural based fragment growth using iMiner.} a). 2D and 3D rendering of the core fragment to grow new molecules from in the Mpro canonical binding pockets, interacting with His163 and Glu166. b). Surface rendering of the filtered set of iMiner generated molecules overlaid in the Mpro canonical binding pockets denoted as P1, P1' \& P2. The substructure fragments are shown in magenta. c). List of the 12 filtered molecules with the fragment highlighted in pink. }
\label{fig:grow}
\end{figure*}

\subsection{Structure-based fragment growth}
Fragment-Based Drug Discovery (FBDD) has emerged as an attractive strategy in modern medicinal chemistry and drug development due to its advantages in lower experimental cost and diversity of paths to novel compounds\cite{erlanson2004}. Fragments, for instance from a screening campaign or as an interesting substructure of a validated binder, can serve as foundational building blocks which can be evolved and optimized into potent drug candidates. While fragments provide only starting points due to their modest affinity, deciding how to grow the molecule or even connect multiple fragments while enhancing activity can be an intricate and subtle task. While FBDD can be readily addressed by iMiner using structural similarity scores as demonstrated above, it is now more challenging because we are also enforcing the structural binding mode of the starting fragment. One intuitive approach would be to perform docking with constraints. However, few publicly available docking programs offer such function without extensive reparametrization. Instead, iMiner employs a "top-down" approach that in addition to rewarding molecules with high similarity, penalizes the docking scores of molecules whose binding poses mismatch the starting fragment, thereby reducing sampling of molecules with incorrect orientation or connectivity during the RL phase. 

As an example we use iMiner to propose molecules containing a pyridinone ring connected to a pyridinyl group that resides in the Mpro P1 pocket (Fig.\ref{fig:grow}a), extracted from the molecules developed by Zhang \textit{et al.} for Mpro, by performing RL training with a pose adjusted Vina score and fragment similarity score starting from a fine-tuned model (see Methods section). A set of 12 molecules in Figure \ref{fig:grow}b-c with the matching substructure aligned to the starting fragment were harvested from the validation and post-filtering process. As seen in the Figure many of the proposed molecules extended from the fragment adopt a cloverleaf-like motif branching into the P1, P1' and P2 subpockets as observed in Zhang \textit{et al.}. The preservation of hydrogen bonding interactions to His163 and Glu166 of these predicted poses as adopted by the original fragment were also preserved in the iMiner molecules. Together the results verify that the RL-physics model learned not only the chemical and structural information of the fragment,  but was able to grow a larger molecule  that improves the drug complementarity to binding sites. 

\begin{figure*}[tbp]
\centering
\includegraphics[width=0.92\textwidth]{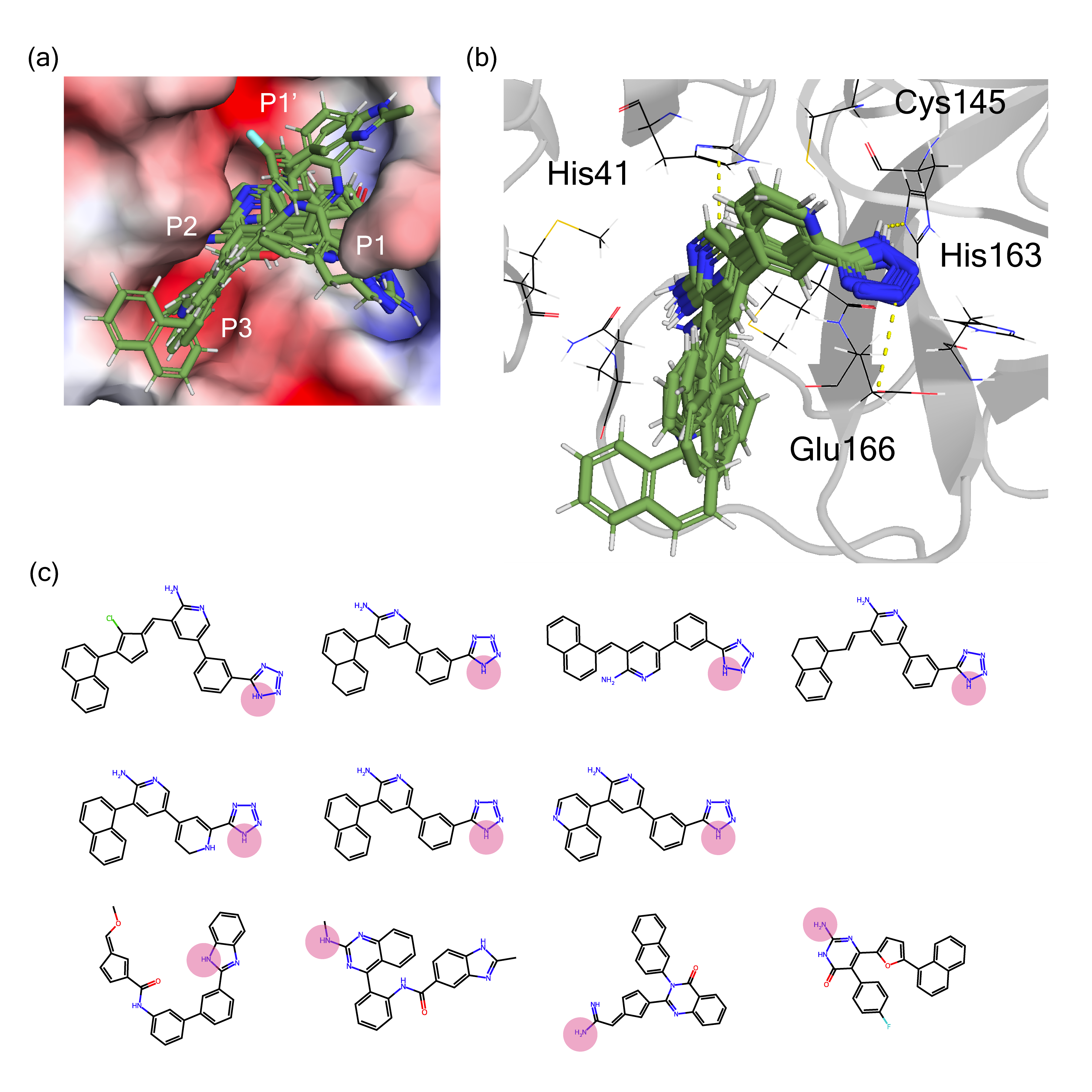}
\caption{\textit{The filtered set of iMiner molecules designed to interact with Mpro residue His163.} a). Surface rendering of the generated molecules in the Mpro canonical binding pockets. b). Binding modes of the generated molecules that shares a common tretrazole motif. c). List of the filtered set of iMiner molecules with hydrogen bond donors to His163 highlighted in pink.}
\label{fig:interaction}
\end{figure*}

\subsection{Designing molecules to interact with specific protein sites}
An alternative scenario of significant interest involves designing molecules to engage with specific protein residues within the pocket. These residues may possess vital catalytic activities or exhibit low mutation rates, thus the inhibitors designed specific to these residues are less likely to develop resistance to viral variants. We investigated the use of iMiner to generate molecules with hydrogen bonding interaction to His163 in the SARS-CoV-2 active site. His163 is a conserved residue in different SARS-CoV-2 strains and multiple inhibitors such as the FDA-approved Remdesivir\cite{Beigel2020} and Ensitrelvir\cite{Unoh2022}, show consensus in forming a hydrogen bond with His163. Starting from the ChEMBL pretrained model and assuming no other knowledge about the structural and chemical features of a potential Mpro binder, we trained a model using an interaction score with respect to His163, Vina docking score and drug-likeliness score in the reward function. 

In Figure \ref{fig:interaction}, the overlay of post-filtered iMiner generated molecules in their binding conformations determined through AutoDock Vina confirms their participation in hydrogen bonding interaction with His163 as expected. The set of molecules can be clustered into two groups: one that presents azole derivatives as a result of the interaction driven learning (Fig.\ref{fig:interaction}c top 2 rows), and the rest that adopts a L-shaped conformation around the P1, P2 and P1' subpockets (Fig.\ref{fig:interaction}c bottom row). Figure \ref{fig:interaction}b also shows that the generated molecules containing tetrazole make additional hydrogen bonds with Glu166 and $\pi$ stacking interactions with the catalytic dyad residue His41, although not explicitly enforced by the reward function.

\subsection{Binding stability analysis} 
In addition to Lipinski rule violations and elimination of PAINS, a final step in the workflow evaluation of generated molecules using the iMiner algorithm is to select molecules based on the binding affinities of the new scoring functions. This is assessed through a consensus of molecules ranked with newVina and newIGN scoring functions\cite{li2021computational}, and subsequently we perform long MD simulations to measure stability of the consensus molecules. 

To illustrate the stability workflow we analyze the unconditional \textit{de novo} design set. We selected the molecules that rank both in the top 300 of newVina rescoring and newIGN rescoring, resulting in an overlap of 29 molecules in total. In each case we start with 50-ns MD simulations, of which we found that 7 molecules had robust binding stability as defined by an average RMSD smaller than 4 $\mathrm{\mathring{A}}$ and preserving more than 1 hydrogen bond between the ligand and protein residues. In order to validate the MD results for these 7 molecules more rigorously, the simulation was then prolonged to 1 $\mathrm{\mu s}$, and we found that 3 generated molecules retained good binding within the active site as measured by the RMSD relative to the original binding pose (Figure S3). For comparison, we also performed 1 $\mathrm{\mu s}$ simulation for two known binders (PDB code: 7L11 and 7L13) reported by Zhang \textit{et al.}\cite{zhang2021potent}.

\begin{figure*}
    \centering
    \includegraphics[width=0.92\textwidth]{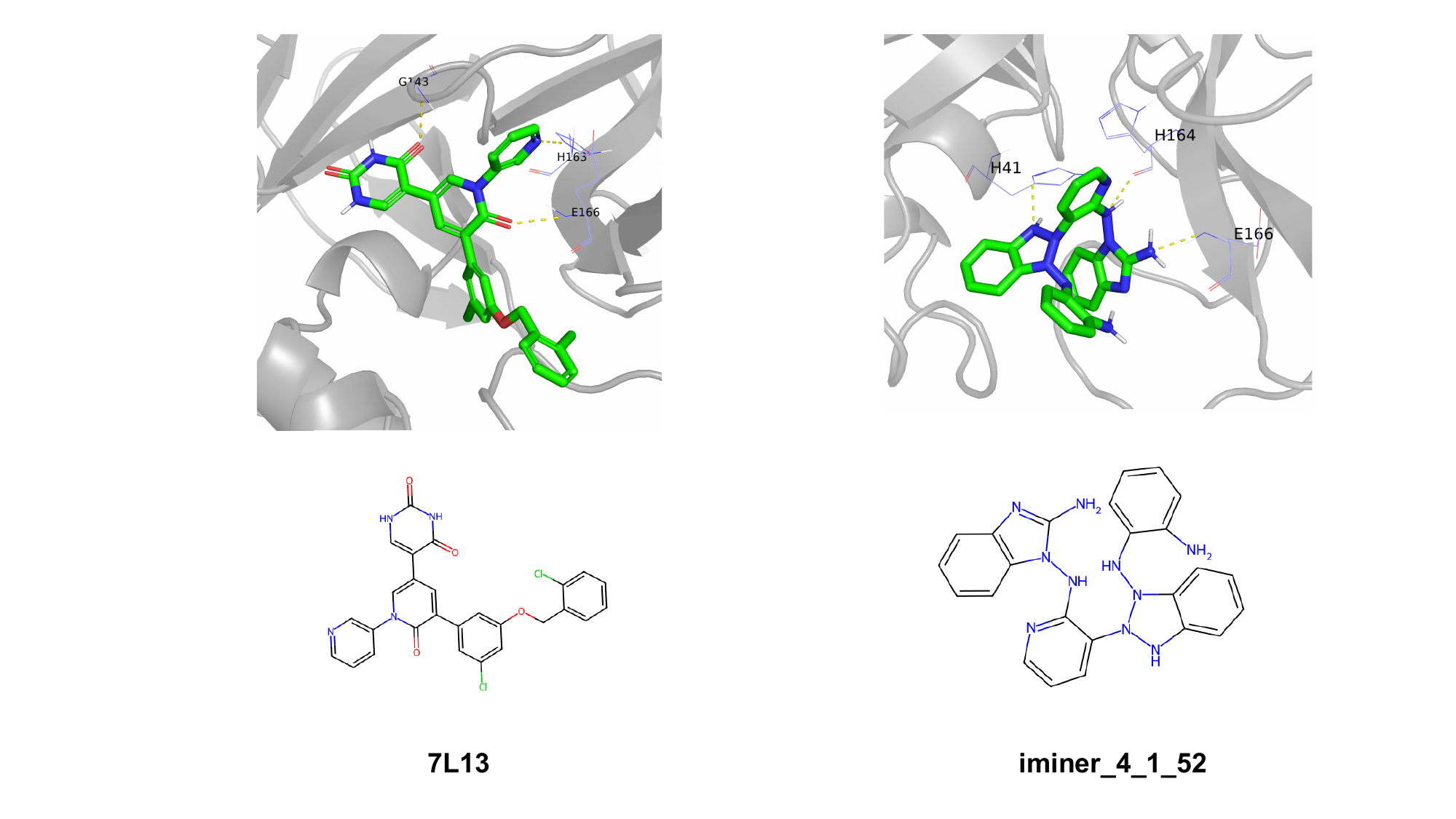}
    \includegraphics[width=0.92\textwidth]{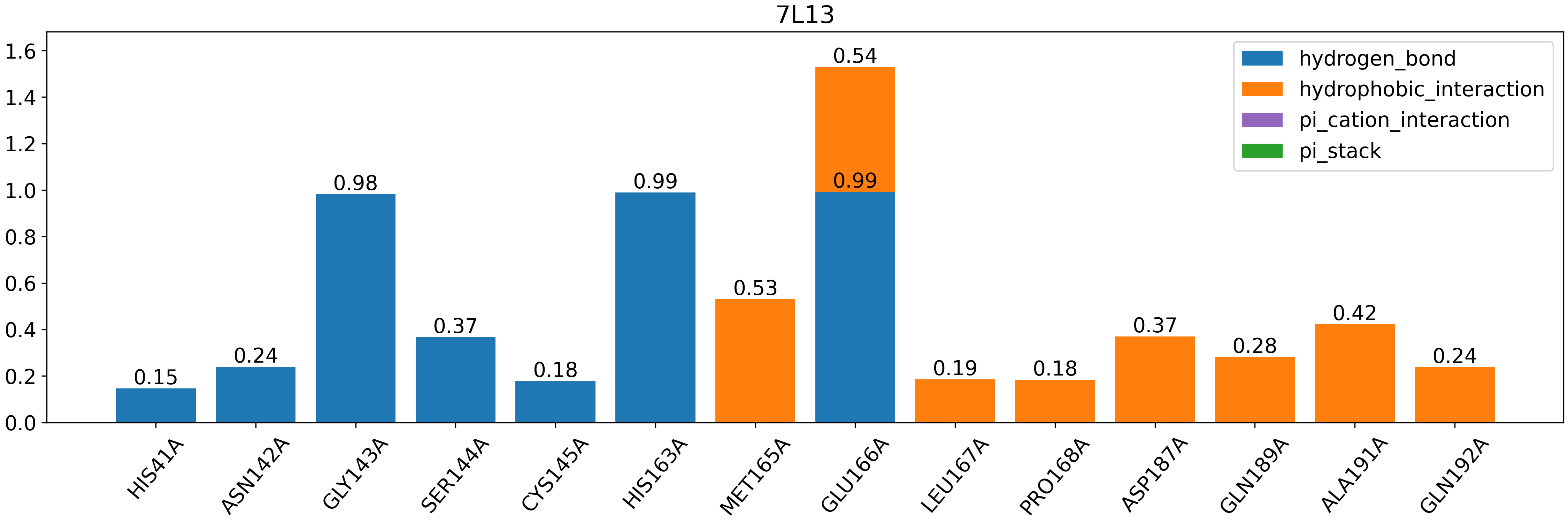}
    \includegraphics[width=0.92\textwidth]{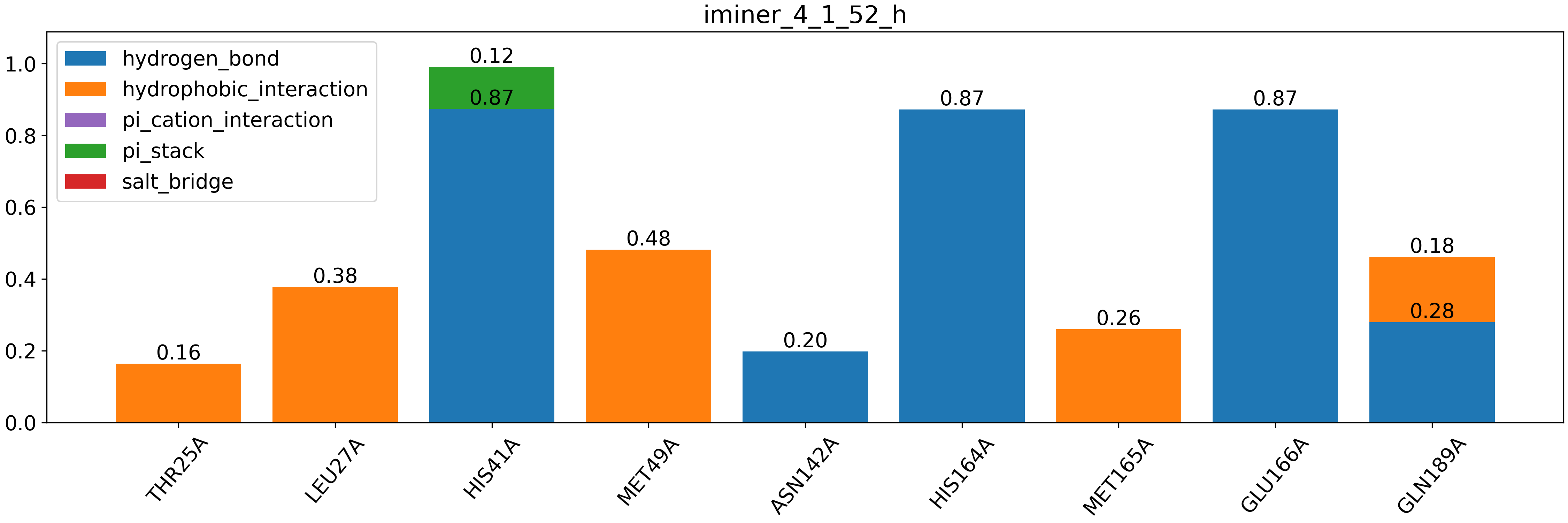}
    \caption{\textit{iMiner workflow for post-analysis.} (top) Representative 2D and 3D structures of molecules 7L13 from Zhang \textit{et al.}\cite{zhang2021potent} and iMiner 4-1-52 from the unconditional \textit{de novo} design. (bottom) A comparison of the molecular interactions between ligand and protein and their frequency over 1$\mu$s MD simulation for 7L13 and iMiner 4-1-52.}
    \label{fig:task1_md}
\end{figure*}

As is evident in Figure 7 and Figure S4, the iMiner generated molecules form very stable interactions such as hydrogen bonds, hydrophobic interactions, and pi-stacking, just as is the case for the known binders 7L11 and 7L13.\cite{zhang2021potent} In Figure 7 we find that both 7L13 and iMiner 4-1-52 form hydrogen bonds with Glu166, which is desirable as it is a low mutation site residue in Mpro such that these molecules may remain viable against future variants of SARS-Cov-2\cite{hegyi2002conservation,Moghadasi2023}. The iMiner molecule forms a long-lived hydrogen bond with His164 vs. His163 utilized by 7L13, and a stable hydrogen bond with His41 which is only weakly populated in 7L13. Similar conclusions are drawn when comparing the known binders against iMiner 4-1-1 and iMiner 4-1-250. Thus at the end of the iMiner workflow, the 3 generated molecules are certainly worth pursuing for synthesis and validation against biochemical and cellular assays.

\section{Conclusions}
\label{sec:conclusion}
\noindent
In this work we have shown that by combining real-time docking of 3D structures with state-of-the-art reinforcement learning algorithms, we can efficiently navigate through uncharted regions of chemical space while maintaining good metrics for synthetic feasibility and drug-likeness. The flexibility of the iMiner physics-ML model also allows for creation of  molecules that enforce interactions with target active site residues, as well as growing molecules from fragments with options for satisfying chemical or structural restraints. As illustrated using the exemplar target, the Mpro catalytic site, the generated inhibitor molecules proposed by iMiner are optimized with respect to shape and intermolecular interactions to the target protein, but are also diverse enough when compared to other predicted Mpro inhibitor datasets, i.e. molecules submitted to the COVID-moonshot project\cite{boby2023open} or the trefoil inhibitors optimized by the Jorgensen group.\cite{zhang2021potent} Finally, every aspect of this work is generalizable to other protein targets and beyond the active site, for example allosteric inhibitora. 

Overall, we believe the iMiner RL-physics algorithm and workflow tool will be of great benefit to the computational and medicinal chemistry fields at large, and potentially aid traditional drug-design workflows as well. Although we have focused our current work on targeting Mpro of SARS-CoV-2, extension of this work to other protein targets relevant to other global diseases would be relatively trivial. For example,  bacterial resistance to antibiotics is of preeminent concern in the medical community\cite{Ventola2015}, and our iMiner workflow approach could be used to target novel bacterial biomolecules, such as bacterial Ribosomes, or target resistance conferring bacterial proteins such as $\beta$-lactamase.\cite{Ventola2015} Another direction pertain to molecules that are experimentally validated through a traditional HTVS approach as good binders, in which the iMiner algorithm could be utilized as an optimization or refinement step for elaborating on these existing leads or scaffolds. The potential of the method in this direction will be explored in future work.

\section{Methods}
\label{sec:methods}
\noindent
\textbf{Neural network architecture}. The generative model employed in this study is an ASGD Weight-Dropped LSTM (AWD-LSTM)\cite{Merity2018}, which is a specific variant of the Long Short Term Memory (LSTM) recurrent neural network with shared DropConnect for weight regularization, and was trained through a non-monotonically triggered average stochastic gradient descent (NT-ASGD) algorithm.\cite{Merity2018,polyak1992acceleration} The basic LSTM cell contains two internal states, the hidden state $h_t$ and the cell state $c_t$, and can be described through the following set of equations:

\begin{align}
    i_t&=\sigma(W^ix_t+U^ih_{t-1}) \\
    f_t&=\sigma(W^fx_t+U^fh_{t-1})  \\
    o_t&=\sigma(W^ox_t+U^oh_{t-1}) \\
    \Tilde{c}_t&=\tanh{(W^cx_t+U^ch_{t-1})} \\
    c_t&=i_t\odot\Tilde{c}_t+f_t\odot c_{t-1} \\
    h_t&=o_t\odot\tanh{c_t}
\end{align}
\noindent
where $[W^i,W^f,W^o,W^c,U^i,U^f,U^o,U^c]$ are the trainable parameters of the model, $x_t$ is the input to the cell at the current timestep, $\Tilde{c}_t$ contains the information to be added to the cell state, and $i_t,f_t,o_t$ represent the update gate, forget gate and output gate respectively, which are numbers between $(0,1)$ that controls how much information should be updated, discarded or retrieved from the cell state. $\sigma$ denotes the sigmoid function, and $\odot$ represents element-wise multiplication. The DropConnect mechanism\cite{wan2013regularization} was applied to the hidden-to-hidden weight matrices $[U^i,U^f,U^o,U^c]$ by randomly zeroing out a small portion of the parameters in these weight matrices to prevent overfitting and ensured that the same positions in the hidden vectors were treated consistently throughout the forward and backward pass in regards to whether or not to be dropped.

The inputs into the RNN cells were tokens embedded as vectors of length 400, and 3 LSTM cells were stacked sequentially, that had 1152, 1152 and 400 units each. The hidden state from the last timestep of the last RNN cell was then connected to a linear decoder with output size of 56 and softmax activation, representing the probabilities of the 56 possible tokens from the vocabulary. The dropout values used in the model were: embedding dropout=0.002, LSTM weight dropout=0.02, RNN hidden state dropout=0.015 and output dropout=0.01.
The neural network was implemented using pyTorch\cite{paszke2017automatic} and the fastai package\cite{howard2020fastai}.

\textbf{Supervised pretraining of the network}
The generative model was pretrained using molecules from ChEMBL 24\cite{Gaulton2012}, and a total of 1,440,263 molecules were selected for training. All molecules were first converted to SELFIES strings using the selfies python package\cite{Krenn2020}, and the tokens were extracted from the SELFIES strings with fastai language model. We used categorical cross entropy loss:
\begin{align}
  L_{\Theta}=-\frac{1}{N}\sum_{i=1}^N\sum_{t_i}\hat{p}(t_i|t_1, t_2,...,t_{i-1}) \log{p_{\Theta}(t_i|t_1, t_2,...,t_{i-1})}  
\end{align}
 where N represents the number of tokens in a molecule, $\hat{p}(t_i|t_1, t_2,...,t_{i-1})$ represents the actual probability of a specific token in the string at position $i$ and with all previous defined tokens $t_1$ through $t_{i-1}$, and ${p_{\Theta}(t_i|t_1, t_2,...,t_{i-1})}$ the probability predicted by the neural network with parameters $\Theta$. The model was trained using Adam optimizer\cite{kingma2014adam} in batches of size 512, and we employed the ``one cycle" learning rate policy\cite{smith2018disciplined} with the maximum learning rate of 0.0005 to achieve superconvergence\cite{smith2019super}. During this pretraining stage we also used weight decay=0.01 and the dropout multiplier of 0.2. The model was pretrained for 30 epochs.
 
For iMiner used to generate molecules similar to a scaffold or growing from a fragment, a fine-tuning was conducted using molecules containing the target scaffold or fragment. These molecules can be curated from open source databases via substructure search, or from a set of molecules of any size defined by the users. While it is possible to directly perform RL training from the ChEMBL pretrained model with a fragment similarity reward term, the additional training ensures a consistent model performance in finding molecules with the desired substructures, as otherwise some specific fragments can be rare in the ChEMBL24 dataset and the model cannot make sufficient sampling of the relevant structures in the initial few epochs of RL training. Details of the additional fine-tuning are provided in the Supplementary Text.
 
\textbf{Reinforcement learning procedure}. Our RL training target goal is to maximize $J(\Theta)$ from formula(\ref{formula:rl_target}) by taking steps along $\partial_{\Theta}J(\Theta)$.
The exact value for $J(\Theta)$ is intractable to evaluate, but can be approximated through sampling the distribution of $s_T$, which gives
\begin{align}
    J(\Theta)\approx\sum_{S_T}p_{\Theta}(s_T) r(s_T)
\end{align}
and then
\begin{align}
    \partial_{\Theta}J(\Theta) &=\sum_{s_T}[\partial_{\Theta}p_{\Theta}(s_T)] r(s_T) \\
    &=\sum_{s_T}p_{\Theta}(s_T)[\sum_{t=1}^T \partial_{\Theta}\log p_{\Theta}(s_t|s_{t-1})] r(s_T)
    \label{formula:reinforce}
\end{align}
Directly taking gradients according to (\ref{formula:reinforce}) corresponds to the REINFORCE algorithm\cite{williams1992simple}. In this work we further utilized the PPO algorithm\cite{schulman2017proximal}, which estimated the gradients through a clipped reward and with an extra entropy bonus term:
\begin{align}
    J'(\Theta)=\sum_{s_T}p_{\Theta}(s_T)[\sum_{t=1}^T J_t^{\mathrm{CLIP}}(\Theta)+\alpha S[p_{\Theta}(s_t|s_{t-1})]]
    \label{formula:ppo}
\end{align}
where 
\begin{align}
    J_t^{\mathrm{CLIP}}(\Theta)=\min(R_t(\Theta) r(s_T), \mathrm{clip}(R_t(\Theta), 1-\epsilon, 1+\epsilon) r(s_T))
\end{align}
with
\begin{align}
        R_t(\Theta)&=\frac{p_{\Theta}(s_t|s_{t-1})}{p_{\Theta_{\mathrm{old}}}(s_t|s_{t-1})}
\end{align}
denoting the ratio between the probability distribution in the current iteration and the probability distribution from the previous iteration (the iteration before last gradient update). A PPO algorithm reduces variance in the gradient, stabilizes training runs, and also encourages the model to explore a wider region of the chemical space through the introduction of an entropy bonus term. 

After the pretraining finished, we copied the weights to a separate model with identical architecture and trained with reinforcement learning using PPO. Because AutoDock Vina can predict different scores and poses for the same molecules due to random initiation, docking-derived reward terms for the same molecules sampled multiple times in one batch were averaged before the target functions were calculated for training stability. In each iteration after the molecules were sampled, model weights were updated by taking gradient steps on the target function through formula (\ref{formula:ppo}) using Adam optimizer. In each iteration, all collected data were used for training the model for a maximum of 10 epochs. The trainer would continue into next iteration and collect new molecules for training if the K-L divergence between the latest predicted probability and the old probability exceeded 0.03. 

The model was trained with RL until the mean entropy of the predicted probability of the tokens from the RNN started to decrease drastically. The training details including batch size, learning rate, hyperparameters $\alpha$ and $\epsilon$ and the ratio between the reward terms $\lambda$ for the 4 tasks are report in Table S2.

\begin{acknowledgement}
This work was supported by National Institute of Allergy and Infectious Disease grant U19-AI171954. This research used computational resources of the National Energy Research Scientific Computing, a DOE Office of Science User Facility supported by the Office of Science of the U.S. Department of Energy under Contract No. DE-AC02-05CH11231. 
\end{acknowledgement}

\section{Availability and Implementation}
All the code for the iMiner reinforcement learning algorithm and workflow are provided in a private GitHub repository: https://github.com/THGLab/iMiner under reaosonable requests.

\begin{suppinfo}
Docking preparation and procedures, analysis of the generated molecules and additional model information are provided in the Supporting Information. \textcolor{red}{finish SI.}
\end{suppinfo}

\section{Author Contributions}
\noindent
J.L., O. Z. and T.H-G. conceived the scientific direction, designed the experiments, analyzed results, and wrote the  manuscript. F.L.K. prepared Mpro structures for docking in AutoDock Vina. C.P. wrote the code for pre-training the network. O.Z. wrote the code for the AutoDock Vina workflow. Y.W. wrote codes for molecular dynamics simulations and subsequent analysis. D.B., X.G., and K.S. contributed ideas and suggestions to the work. R.E.A. provided drug-design  guidance. All authors discussed and performed final edits to the manuscript.

\section*{Declaration of Interests}
\noindent
C.P. has equity interest in Athae Bio. The other authors declare no competing interests.

\bibliography{references}

\providecommand{\latin}[1]{#1}
\makeatletter
\providecommand{\doi}
  {\begingroup\let\do\@makeother\dospecials
  \catcode`\{=1 \catcode`\}=2 \doi@aux}
\providecommand{\doi@aux}[1]{\endgroup\texttt{#1}}
\makeatother
\providecommand*\mcitethebibliography{\thebibliography}
\csname @ifundefined\endcsname{endmcitethebibliography}  {\let\endmcitethebibliography\endthebibliography}{}
\begin{mcitethebibliography}{68}
\providecommand*\natexlab[1]{#1}
\providecommand*\mciteSetBstSublistMode[1]{}
\providecommand*\mciteSetBstMaxWidthForm[2]{}
\providecommand*\mciteBstWouldAddEndPuncttrue
  {\def\EndOfBibitem{\unskip.}}
\providecommand*\mciteBstWouldAddEndPunctfalse
  {\let\EndOfBibitem\relax}
\providecommand*\mciteSetBstMidEndSepPunct[3]{}
\providecommand*\mciteSetBstSublistLabelBeginEnd[3]{}
\providecommand*\EndOfBibitem{}
\mciteSetBstSublistMode{f}
\mciteSetBstMaxWidthForm{subitem}{(\alph{mcitesubitemcount})}
\mciteSetBstSublistLabelBeginEnd
  {\mcitemaxwidthsubitemform\space}
  {\relax}
  {\relax}

\bibitem[Neves \latin{et~al.}(2018)Neves, Braga, Melo-Filho, Moreira-Filho, Muratov, and Andrade]{Neves2018}
Neves,~B.~J.; Braga,~R.~C.; Melo-Filho,~C.~C.; Moreira-Filho,~J.~T.; Muratov,~E.~N.; Andrade,~C.~H. QSAR-Based Virtual Screening: Advances and Applications in Drug Discovery. \emph{Front. Pharma.} \textbf{2018}, \emph{9}\relax
\mciteBstWouldAddEndPuncttrue
\mciteSetBstMidEndSepPunct{\mcitedefaultmidpunct}
{\mcitedefaultendpunct}{\mcitedefaultseppunct}\relax
\EndOfBibitem
\bibitem[Batool \latin{et~al.}(2019)Batool, Ahmad, and Choi]{Batool2019}
Batool,~M.; Ahmad,~B.; Choi,~S. A Structure-Based Drug Discovery Paradigm. \emph{Inter. J. Mol. Sci.} \textbf{2019}, \emph{20}\relax
\mciteBstWouldAddEndPuncttrue
\mciteSetBstMidEndSepPunct{\mcitedefaultmidpunct}
{\mcitedefaultendpunct}{\mcitedefaultseppunct}\relax
\EndOfBibitem
\bibitem[Meng \latin{et~al.}(2011)Meng, Zhang, Mezei, and Cui]{Meng2011}
Meng,~X.~Y.; Zhang,~H.~X.; Mezei,~M.; Cui,~M. Molecular docking: a powerful approach for structure-based drug discovery. \emph{Curr. Comput. Aided Drug Des.} \textbf{2011}, \emph{7}, 146--57\relax
\mciteBstWouldAddEndPuncttrue
\mciteSetBstMidEndSepPunct{\mcitedefaultmidpunct}
{\mcitedefaultendpunct}{\mcitedefaultseppunct}\relax
\EndOfBibitem
\bibitem[Amaro and Mulholland(2018)Amaro, and Mulholland]{Amaro2018}
Amaro,~R.~E.; Mulholland,~A.~J. Multiscale methods in drug design bridge chemical and biological complexity in the search for cures. \emph{Nature Rev. Chem.} \textbf{2018}, \emph{2}, 0148\relax
\mciteBstWouldAddEndPuncttrue
\mciteSetBstMidEndSepPunct{\mcitedefaultmidpunct}
{\mcitedefaultendpunct}{\mcitedefaultseppunct}\relax
\EndOfBibitem
\bibitem[Gaulton \latin{et~al.}(2012)Gaulton, Bellis, Bento, Chambers, Davies, Hersey, Light, McGlinchey, Michalovich, Al-Lazikani, and Overington]{Gaulton2012}
Gaulton,~A.; Bellis,~L.~J.; Bento,~A.~P.; Chambers,~J.; Davies,~M.; Hersey,~A.; Light,~Y.; McGlinchey,~S.; Michalovich,~D.; Al-Lazikani,~B.; Overington,~J.~P. ChEMBL: a large-scale bioactivity database for drug discovery. \emph{Nuc. Acids Res.} \textbf{2012}, \emph{40}, D1100--D1107\relax
\mciteBstWouldAddEndPuncttrue
\mciteSetBstMidEndSepPunct{\mcitedefaultmidpunct}
{\mcitedefaultendpunct}{\mcitedefaultseppunct}\relax
\EndOfBibitem
\bibitem[Kim \latin{et~al.}(2021)Kim, Chen, Cheng, Gindulyte, He, He, Li, Shoemaker, Thiessen, Yu, Zaslavsky, Zhang, and Bolton]{Kim2021}
Kim,~S.; Chen,~J.; Cheng,~T.; Gindulyte,~A.; He,~J.; He,~S.; Li,~Q.; Shoemaker,~B.~A.; Thiessen,~P.~A.; Yu,~B.; Zaslavsky,~L.; Zhang,~J.; Bolton,~E.~E. PubChem in 2021: new data content and improved web interfaces. \emph{Nuc. Acids Res.} \textbf{2021}, \emph{49}, D1388--D1395\relax
\mciteBstWouldAddEndPuncttrue
\mciteSetBstMidEndSepPunct{\mcitedefaultmidpunct}
{\mcitedefaultendpunct}{\mcitedefaultseppunct}\relax
\EndOfBibitem
\bibitem[Sterling and Irwin(2015)Sterling, and Irwin]{Sterling2015}
Sterling,~T.; Irwin,~J.~J. ZINC 15 – Ligand Discovery for Everyone. \emph{J. Chem. Inform. Model.} \textbf{2015}, \emph{55}, 2324--2337\relax
\mciteBstWouldAddEndPuncttrue
\mciteSetBstMidEndSepPunct{\mcitedefaultmidpunct}
{\mcitedefaultendpunct}{\mcitedefaultseppunct}\relax
\EndOfBibitem
\bibitem[Grygorenko \latin{et~al.}(2020)Grygorenko, Radchenko, Dziuba, Chuprina, Gubina, and Moroz]{Grygorenko2020}
Grygorenko,~O.~O.; Radchenko,~D.~S.; Dziuba,~I.; Chuprina,~A.; Gubina,~K.~E.; Moroz,~Y.~S. Generating Multibillion Chemical Space of Readily Accessible Screening Compounds. \emph{iSci.} \textbf{2020}, \emph{23}\relax
\mciteBstWouldAddEndPuncttrue
\mciteSetBstMidEndSepPunct{\mcitedefaultmidpunct}
{\mcitedefaultendpunct}{\mcitedefaultseppunct}\relax
\EndOfBibitem
\bibitem[Polishchuk \latin{et~al.}(2013)Polishchuk, Madzhidov, and Varnek]{polishchuk2013estimation}
Polishchuk,~P.~G.; Madzhidov,~T.~I.; Varnek,~A. Estimation of the size of drug-like chemical space based on GDB-17 data. \emph{J. Comp.-Aid, Mol. Des.} \textbf{2013}, \emph{27}, 675--679\relax
\mciteBstWouldAddEndPuncttrue
\mciteSetBstMidEndSepPunct{\mcitedefaultmidpunct}
{\mcitedefaultendpunct}{\mcitedefaultseppunct}\relax
\EndOfBibitem
\bibitem[Reulecke \latin{et~al.}(2008)Reulecke, Lange, Albrecht, Klein, and Rarey]{reulecke2008towards}
Reulecke,~I.; Lange,~G.; Albrecht,~J.; Klein,~R.; Rarey,~M. Towards an integrated description of hydrogen bonding and dehydration: decreasing false positives in virtual screening with the HYDE scoring function. \emph{ChemMedChem: Chemistry Enabling Drug Discovery} \textbf{2008}, \emph{3}, 885--897\relax
\mciteBstWouldAddEndPuncttrue
\mciteSetBstMidEndSepPunct{\mcitedefaultmidpunct}
{\mcitedefaultendpunct}{\mcitedefaultseppunct}\relax
\EndOfBibitem
\bibitem[Duffy \latin{et~al.}(2012)Duffy, Zhu, Decornez, and Kitchen]{duffy2012early}
Duffy,~B.~C.; Zhu,~L.; Decornez,~H.; Kitchen,~D.~B. Early phase drug discovery: cheminformatics and computational techniques in identifying lead series. \emph{Bioorg. Med. Chem.} \textbf{2012}, \emph{20}, 5324--5342\relax
\mciteBstWouldAddEndPuncttrue
\mciteSetBstMidEndSepPunct{\mcitedefaultmidpunct}
{\mcitedefaultendpunct}{\mcitedefaultseppunct}\relax
\EndOfBibitem
\bibitem[Sanchez-Lengeling and Aspuru-Guzik(2018)Sanchez-Lengeling, and Aspuru-Guzik]{sanchez2018inverse}
Sanchez-Lengeling,~B.; Aspuru-Guzik,~A. Inverse molecular design using machine learning: Generative models for matter engineering. \emph{Science} \textbf{2018}, \emph{361}, 360--365\relax
\mciteBstWouldAddEndPuncttrue
\mciteSetBstMidEndSepPunct{\mcitedefaultmidpunct}
{\mcitedefaultendpunct}{\mcitedefaultseppunct}\relax
\EndOfBibitem
\bibitem[Kusner \latin{et~al.}(2017)Kusner, Paige, and Hern{\'a}ndez-Lobato]{kusner2017grammar}
Kusner,~M.~J.; Paige,~B.; Hern{\'a}ndez-Lobato,~J.~M. Grammar variational autoencoder. International Conference on Machine Learning. 2017; pp 1945--1954\relax
\mciteBstWouldAddEndPuncttrue
\mciteSetBstMidEndSepPunct{\mcitedefaultmidpunct}
{\mcitedefaultendpunct}{\mcitedefaultseppunct}\relax
\EndOfBibitem
\bibitem[Dai \latin{et~al.}(2018)Dai, Tian, Dai, Skiena, and Song]{dai2018syntax}
Dai,~H.; Tian,~Y.; Dai,~B.; Skiena,~S.; Song,~L. Syntax-directed variational autoencoder for structured data. \emph{arXiv preprint arXiv:1802.08786} \textbf{2018}, \relax
\mciteBstWouldAddEndPunctfalse
\mciteSetBstMidEndSepPunct{\mcitedefaultmidpunct}
{}{\mcitedefaultseppunct}\relax
\EndOfBibitem
\bibitem[Subramanian \latin{et~al.}(2020)Subramanian, Saha, Sharma, Tailor, and Satapathi]{subramanian2020inverse}
Subramanian,~A.; Saha,~U.; Sharma,~T.; Tailor,~N.~K.; Satapathi,~S. Inverse Design of Potential Singlet Fission Molecules using a Transfer Learning Based Approach. \emph{arXiv preprint arXiv:2003.07666} \textbf{2020}, \relax
\mciteBstWouldAddEndPunctfalse
\mciteSetBstMidEndSepPunct{\mcitedefaultmidpunct}
{}{\mcitedefaultseppunct}\relax
\EndOfBibitem
\bibitem[Olivecrona \latin{et~al.}(2017)Olivecrona, Blaschke, Engkvist, and Chen]{olivecrona2017molecular}
Olivecrona,~M.; Blaschke,~T.; Engkvist,~O.; Chen,~H. Molecular de-novo design through deep reinforcement learning. \emph{J. Cheminform.} \textbf{2017}, \emph{9}, 1--14\relax
\mciteBstWouldAddEndPuncttrue
\mciteSetBstMidEndSepPunct{\mcitedefaultmidpunct}
{\mcitedefaultendpunct}{\mcitedefaultseppunct}\relax
\EndOfBibitem
\bibitem[Popova \latin{et~al.}(2018)Popova, Isayev, and Tropsha]{popova2018deep}
Popova,~M.; Isayev,~O.; Tropsha,~A. Deep reinforcement learning for de novo drug design. \emph{Sci. Adv.} \textbf{2018}, \emph{4}, eaap7885\relax
\mciteBstWouldAddEndPuncttrue
\mciteSetBstMidEndSepPunct{\mcitedefaultmidpunct}
{\mcitedefaultendpunct}{\mcitedefaultseppunct}\relax
\EndOfBibitem
\bibitem[Gottipati \latin{et~al.}(2020)Gottipati, Sattarov, Niu, Pathak, Wei, Liu, Blackburn, Thomas, Coley, Tang, \latin{et~al.} others]{gottipati2020learning}
Gottipati,~S.~K.; Sattarov,~B.; Niu,~S.; Pathak,~Y.; Wei,~H.; Liu,~S.; Blackburn,~S.; Thomas,~K.; Coley,~C.; Tang,~J.; others Learning to navigate the synthetically accessible chemical space using reinforcement learning. International Conference on Machine Learning. 2020; pp 3668--3679\relax
\mciteBstWouldAddEndPuncttrue
\mciteSetBstMidEndSepPunct{\mcitedefaultmidpunct}
{\mcitedefaultendpunct}{\mcitedefaultseppunct}\relax
\EndOfBibitem
\bibitem[Zhavoronkov \latin{et~al.}(2019)Zhavoronkov, Ivanenkov, Aliper, Veselov, Aladinskiy, Aladinskaya, Terentiev, Polykovskiy, Kuznetsov, Asadulaev, \latin{et~al.} others]{zhavoronkov2019deep}
Zhavoronkov,~A.; Ivanenkov,~Y.~A.; Aliper,~A.; Veselov,~M.~S.; Aladinskiy,~V.~A.; Aladinskaya,~A.~V.; Terentiev,~V.~A.; Polykovskiy,~D.~A.; Kuznetsov,~M.~D.; Asadulaev,~A.; others Deep learning enables rapid identification of potent DDR1 kinase inhibitors. \emph{Nature Biotech.} \textbf{2019}, \emph{37}, 1038--1040\relax
\mciteBstWouldAddEndPuncttrue
\mciteSetBstMidEndSepPunct{\mcitedefaultmidpunct}
{\mcitedefaultendpunct}{\mcitedefaultseppunct}\relax
\EndOfBibitem
\bibitem[Bung \latin{et~al.}(2021)Bung, Krishnan, Bulusu, and Roy]{bung2021novo}
Bung,~N.; Krishnan,~S.~R.; Bulusu,~G.; Roy,~A. De novo design of new chemical entities for SARS-CoV-2 using artificial intelligence. \emph{Future Med. Chem.} \textbf{2021}, \emph{13}, 575--585\relax
\mciteBstWouldAddEndPuncttrue
\mciteSetBstMidEndSepPunct{\mcitedefaultmidpunct}
{\mcitedefaultendpunct}{\mcitedefaultseppunct}\relax
\EndOfBibitem
\bibitem[Born \latin{et~al.}(2021)Born, Manica, Cadow, Markert, Mill, Filipavicius, Janakarajan, Cardinale, Laino, and Mart{\'\i}nez]{born2021data}
Born,~J.; Manica,~M.; Cadow,~J.; Markert,~G.; Mill,~N.~A.; Filipavicius,~M.; Janakarajan,~N.; Cardinale,~A.; Laino,~T.; Mart{\'\i}nez,~M.~R. Data-driven molecular design for discovery and synthesis of novel ligands: a case study on SARS-CoV-2. \emph{Mach. Learn.: Sci. Tech.} \textbf{2021}, \emph{2}, 025024\relax
\mciteBstWouldAddEndPuncttrue
\mciteSetBstMidEndSepPunct{\mcitedefaultmidpunct}
{\mcitedefaultendpunct}{\mcitedefaultseppunct}\relax
\EndOfBibitem
\bibitem[Zhang and Chen(2022)Zhang, and Chen]{zhang2022novo}
Zhang,~J.; Chen,~H. De novo molecule design using molecular generative models constrained by ligand--protein interactions. \emph{Journal of Chemical Information and Modeling} \textbf{2022}, \emph{62}, 3291--3306\relax
\mciteBstWouldAddEndPuncttrue
\mciteSetBstMidEndSepPunct{\mcitedefaultmidpunct}
{\mcitedefaultendpunct}{\mcitedefaultseppunct}\relax
\EndOfBibitem
\bibitem[Li \latin{et~al.}(2023)Li, Hu, Ke, Yang, Chen, Xiong, Liu, and Hong]{li2023ls}
Li,~S.; Hu,~C.; Ke,~S.; Yang,~C.; Chen,~J.; Xiong,~Y.; Liu,~H.; Hong,~L. LS-MolGen: Ligand-and-Structure Dual-Driven Deep Reinforcement Learning for Target-Specific Molecular Generation Improves Binding Affinity and Novelty. \emph{Journal of Chemical Information and Modeling} \textbf{2023}, \relax
\mciteBstWouldAddEndPunctfalse
\mciteSetBstMidEndSepPunct{\mcitedefaultmidpunct}
{}{\mcitedefaultseppunct}\relax
\EndOfBibitem
\bibitem[Jeon and Kim(2020)Jeon, and Kim]{jeon2020autonomous}
Jeon,~W.; Kim,~D. Autonomous molecule generation using reinforcement learning and docking to develop potential novel inhibitors. \emph{Sci. Rep.} \textbf{2020}, \emph{10}, 1--11\relax
\mciteBstWouldAddEndPuncttrue
\mciteSetBstMidEndSepPunct{\mcitedefaultmidpunct}
{\mcitedefaultendpunct}{\mcitedefaultseppunct}\relax
\EndOfBibitem
\bibitem[Peng \latin{et~al.}(2022)Peng, Luo, Guan, Xie, Peng, and Ma]{pocket2mol}
Peng,~X.; Luo,~S.; Guan,~J.; Xie,~Q.; Peng,~J.; Ma,~J. Pocket2mol: Efficient molecular sampling based on 3d protein pockets. International Conference on Machine Learning. 2022; pp 17644--17655\relax
\mciteBstWouldAddEndPuncttrue
\mciteSetBstMidEndSepPunct{\mcitedefaultmidpunct}
{\mcitedefaultendpunct}{\mcitedefaultseppunct}\relax
\EndOfBibitem
\bibitem[Zhang \latin{et~al.}(2023)Zhang, Zhang, Jin, Zhang, Hu, Shen, Cao, Du, Kang, Deng, \latin{et~al.} others]{zhang2023resgen}
Zhang,~O.; Zhang,~J.; Jin,~J.; Zhang,~X.; Hu,~R.; Shen,~C.; Cao,~H.; Du,~H.; Kang,~Y.; Deng,~Y.; others ResGen is a pocket-aware 3D molecular generation model based on parallel multiscale modelling. \emph{Nature Machine Intelligence} \textbf{2023}, \emph{5}, 1020--1030\relax
\mciteBstWouldAddEndPuncttrue
\mciteSetBstMidEndSepPunct{\mcitedefaultmidpunct}
{\mcitedefaultendpunct}{\mcitedefaultseppunct}\relax
\EndOfBibitem
\bibitem[Schneuing \latin{et~al.}(2022)Schneuing, Du, Harris, Jamasb, Igashov, Du, Blundell, Li{\'o}, Gomes, Welling, \latin{et~al.} others]{schneuing2022structure}
Schneuing,~A.; Du,~Y.; Harris,~C.; Jamasb,~A.; Igashov,~I.; Du,~W.; Blundell,~T.; Li{\'o},~P.; Gomes,~C.; Welling,~M.; others Structure-based drug design with equivariant diffusion models. \emph{arXiv preprint arXiv:2210.13695} \textbf{2022}, \relax
\mciteBstWouldAddEndPunctfalse
\mciteSetBstMidEndSepPunct{\mcitedefaultmidpunct}
{}{\mcitedefaultseppunct}\relax
\EndOfBibitem
\bibitem[Krenn \latin{et~al.}(2020)Krenn, Häse, Nigam, Friederich, and Aspuru-Guzik]{Krenn2020}
Krenn,~M.; Häse,~F.; Nigam,~A.; Friederich,~P.; Aspuru-Guzik,~A. Self-referencing embedded strings (SELFIES): A 100\% robust molecular string representation. \emph{Mach. Learn.: Sci. Tech.} \textbf{2020}, \emph{1}, 045024\relax
\mciteBstWouldAddEndPuncttrue
\mciteSetBstMidEndSepPunct{\mcitedefaultmidpunct}
{\mcitedefaultendpunct}{\mcitedefaultseppunct}\relax
\EndOfBibitem
\bibitem[Merity \latin{et~al.}(2018)Merity, Keskar, and Socher]{Merity2018}
Merity,~S.; Keskar,~N.~S.; Socher,~R. Regularizing and Optimizing LSTM Language Models. 2018\relax
\mciteBstWouldAddEndPuncttrue
\mciteSetBstMidEndSepPunct{\mcitedefaultmidpunct}
{\mcitedefaultendpunct}{\mcitedefaultseppunct}\relax
\EndOfBibitem
\bibitem[Trott and Olson(2010)Trott, and Olson]{Trott2010}
Trott,~O.; Olson,~A.~J. {AutoDock Vina: Improving the speed and accuracy of docking with a new scoring function, efficient optimization, and multithreading}. \emph{J. Comp. Chem.} \textbf{2010}, \emph{31}, 455--461\relax
\mciteBstWouldAddEndPuncttrue
\mciteSetBstMidEndSepPunct{\mcitedefaultmidpunct}
{\mcitedefaultendpunct}{\mcitedefaultseppunct}\relax
\EndOfBibitem
\bibitem[Lipinski(2000)]{Lipinski2000}
Lipinski,~C.~A. Drug-like properties and the causes of poor solubility and poor permeability. \emph{J. Pharma. Tox. Meth.} \textbf{2000}, \emph{44}, 235--249\relax
\mciteBstWouldAddEndPuncttrue
\mciteSetBstMidEndSepPunct{\mcitedefaultmidpunct}
{\mcitedefaultendpunct}{\mcitedefaultseppunct}\relax
\EndOfBibitem
\bibitem[Dahlin \latin{et~al.}(2015)Dahlin, Nissink, Strasser, Francis, Higgins, Zhou, Zhang, and Walters]{dahlin2015pains}
Dahlin,~J.~L.; Nissink,~J. W.~M.; Strasser,~J.~M.; Francis,~S.; Higgins,~L.; Zhou,~H.; Zhang,~Z.; Walters,~M.~A. PAINS in the assay: chemical mechanisms of assay interference and promiscuous enzymatic inhibition observed during a sulfhydryl-scavenging HTS. \emph{J. Med. Chem.} \textbf{2015}, \emph{58}, 2091--2113\relax
\mciteBstWouldAddEndPuncttrue
\mciteSetBstMidEndSepPunct{\mcitedefaultmidpunct}
{\mcitedefaultendpunct}{\mcitedefaultseppunct}\relax
\EndOfBibitem
\bibitem[Jin \latin{et~al.}(2020)Jin, Du, Xu, Deng, Liu, Zhao, Zhang, Li, Zhang, Peng, \latin{et~al.} others]{jin2020structure}
Jin,~Z.; Du,~X.; Xu,~Y.; Deng,~Y.; Liu,~M.; Zhao,~Y.; Zhang,~B.; Li,~X.; Zhang,~L.; Peng,~C.; others Structure of Mpro from SARS-CoV-2 and discovery of its inhibitors. \emph{Nature} \textbf{2020}, \emph{582}, 289--293\relax
\mciteBstWouldAddEndPuncttrue
\mciteSetBstMidEndSepPunct{\mcitedefaultmidpunct}
{\mcitedefaultendpunct}{\mcitedefaultseppunct}\relax
\EndOfBibitem
\bibitem[Zhang \latin{et~al.}(2021)Zhang, Stone, Deshmukh, Ippolito, Ghahremanpour, Tirado-Rives, Spasov, Zhang, Takeo, Kudalkar, \latin{et~al.} others]{zhang2021potent}
Zhang,~C.-H.; Stone,~E.~A.; Deshmukh,~M.; Ippolito,~J.~A.; Ghahremanpour,~M.~M.; Tirado-Rives,~J.; Spasov,~K.~A.; Zhang,~S.; Takeo,~Y.; Kudalkar,~S.~N.; others Potent noncovalent inhibitors of the main protease of SARS-CoV-2 from molecular sculpting of the drug perampanel guided by free energy perturbation calculations. \emph{ACS Cent. Sci.} \textbf{2021}, \emph{7}, 467--475\relax
\mciteBstWouldAddEndPuncttrue
\mciteSetBstMidEndSepPunct{\mcitedefaultmidpunct}
{\mcitedefaultendpunct}{\mcitedefaultseppunct}\relax
\EndOfBibitem
\bibitem[Cui \latin{et~al.}(2020)Cui, Yang, and Yang]{mpro_review}
Cui,~W.; Yang,~K.; Yang,~H. Recent Progress in the Drug Development Targeting SARS-CoV-2 Main Protease as Treatment for COVID-19. \emph{Front. Mol. Biosci.} \textbf{2020}, \emph{7}, 398\relax
\mciteBstWouldAddEndPuncttrue
\mciteSetBstMidEndSepPunct{\mcitedefaultmidpunct}
{\mcitedefaultendpunct}{\mcitedefaultseppunct}\relax
\EndOfBibitem
\bibitem[Wildman and Crippen(1999)Wildman, and Crippen]{wildman1999prediction}
Wildman,~S.~A.; Crippen,~G.~M. Prediction of physicochemical parameters by atomic contributions. \emph{J. Chem. Inform. Comp. Sci.} \textbf{1999}, \emph{39}, 868--873\relax
\mciteBstWouldAddEndPuncttrue
\mciteSetBstMidEndSepPunct{\mcitedefaultmidpunct}
{\mcitedefaultendpunct}{\mcitedefaultseppunct}\relax
\EndOfBibitem
\bibitem[Brenk \latin{et~al.}(2008)Brenk, Schipani, James, Krasowski, Gilbert, Frearson, and Wyatt]{brenk2008lessons}
Brenk,~R.; Schipani,~A.; James,~D.; Krasowski,~A.; Gilbert,~I.~H.; Frearson,~J.; Wyatt,~P.~G. Lessons learnt from assembling screening libraries for drug discovery for neglected diseases. \emph{ChemMedChem} \textbf{2008}, \emph{3}, 435\relax
\mciteBstWouldAddEndPuncttrue
\mciteSetBstMidEndSepPunct{\mcitedefaultmidpunct}
{\mcitedefaultendpunct}{\mcitedefaultseppunct}\relax
\EndOfBibitem
\bibitem[Tang \latin{et~al.}(2022)Tang, Chen, Lin, Lin, Zhu, Ding, Hu, Ling, and Wu]{vina_gpu}
Tang,~S.; Chen,~R.; Lin,~M.; Lin,~Q.; Zhu,~Y.; Ding,~J.; Hu,~H.; Ling,~M.; Wu,~J. Accelerating autodock vina with gpus. \emph{Molecules} \textbf{2022}, \emph{27}, 3041\relax
\mciteBstWouldAddEndPuncttrue
\mciteSetBstMidEndSepPunct{\mcitedefaultmidpunct}
{\mcitedefaultendpunct}{\mcitedefaultseppunct}\relax
\EndOfBibitem
\bibitem[Schulman \latin{et~al.}(2017)Schulman, Wolski, Dhariwal, Radford, and Klimov]{schulman2017proximal}
Schulman,~J.; Wolski,~F.; Dhariwal,~P.; Radford,~A.; Klimov,~O. Proximal policy optimization algorithms. \emph{arXiv preprint arXiv:1707.06347} \textbf{2017}, \relax
\mciteBstWouldAddEndPunctfalse
\mciteSetBstMidEndSepPunct{\mcitedefaultmidpunct}
{}{\mcitedefaultseppunct}\relax
\EndOfBibitem
\bibitem[Everitt and Hutter(2016)Everitt, and Hutter]{everitt2016avoiding}
Everitt,~T.; Hutter,~M. Avoiding wireheading with value reinforcement learning. International Conference on Artificial General Intelligence. 2016; pp 12--22\relax
\mciteBstWouldAddEndPuncttrue
\mciteSetBstMidEndSepPunct{\mcitedefaultmidpunct}
{\mcitedefaultendpunct}{\mcitedefaultseppunct}\relax
\EndOfBibitem
\bibitem[Bickerton \latin{et~al.}(2012)Bickerton, Paolini, Besnard, Muresan, and Hopkins]{bickerton2012quantifying}
Bickerton,~G.~R.; Paolini,~G.~V.; Besnard,~J.; Muresan,~S.; Hopkins,~A.~L. Quantifying the chemical beauty of drugs. \emph{Nature Chem.} \textbf{2012}, \emph{4}, 90--98\relax
\mciteBstWouldAddEndPuncttrue
\mciteSetBstMidEndSepPunct{\mcitedefaultmidpunct}
{\mcitedefaultendpunct}{\mcitedefaultseppunct}\relax
\EndOfBibitem
\bibitem[Degen \latin{et~al.}(2008)Degen, Wegscheid-Gerlach, Zaliani, and Rarey]{BRICS}
Degen,~J.; Wegscheid-Gerlach,~C.; Zaliani,~A.; Rarey,~M. On the Art of Compiling and Using'Drug-Like'Chemical Fragment Spaces. \emph{ChemMedChem: Chemistry Enabling Drug Discovery} \textbf{2008}, \emph{3}, 1503--1507\relax
\mciteBstWouldAddEndPuncttrue
\mciteSetBstMidEndSepPunct{\mcitedefaultmidpunct}
{\mcitedefaultendpunct}{\mcitedefaultseppunct}\relax
\EndOfBibitem
\bibitem[Gobbi and Poppinger(1998)Gobbi, and Poppinger]{Gobbi1998}
Gobbi,~A.; Poppinger,~D. Genetic optimization of combinatorial libraries. \emph{Biotech. Bioeng.} \textbf{1998}, \emph{61}, 47--54\relax
\mciteBstWouldAddEndPuncttrue
\mciteSetBstMidEndSepPunct{\mcitedefaultmidpunct}
{\mcitedefaultendpunct}{\mcitedefaultseppunct}\relax
\EndOfBibitem
\bibitem[Landrum(2016)]{Landrum2016RDKit2016_09_4}
Landrum,~G. RDKit: Open-Source Cheminformatics Software. \textbf{2016}, \relax
\mciteBstWouldAddEndPunctfalse
\mciteSetBstMidEndSepPunct{\mcitedefaultmidpunct}
{}{\mcitedefaultseppunct}\relax
\EndOfBibitem
\bibitem[Salentin \latin{et~al.}(2015)Salentin, Schreiber, Haupt, Adasme, and Schroeder]{Salentin2015}
Salentin,~S.; Schreiber,~S.; Haupt,~V.~J.; Adasme,~M.~F.; Schroeder,~M. PLIP: fully automated protein-ligand interaction profiler. \emph{Nuc. Acids Res.} \textbf{2015}, \emph{43}, W443--7\relax
\mciteBstWouldAddEndPuncttrue
\mciteSetBstMidEndSepPunct{\mcitedefaultmidpunct}
{\mcitedefaultendpunct}{\mcitedefaultseppunct}\relax
\EndOfBibitem
\bibitem[Lipinski \latin{et~al.}(1997)Lipinski, Lombardo, Dominy, and Feeney]{lipinski1997experimental}
Lipinski,~C.~A.; Lombardo,~F.; Dominy,~B.~W.; Feeney,~P.~J. Experimental and computational approaches to estimate solubility and permeability in drug discovery and development settings. \emph{Adv. Drug Delivery Rev.} \textbf{1997}, \emph{23}, 3--25\relax
\mciteBstWouldAddEndPuncttrue
\mciteSetBstMidEndSepPunct{\mcitedefaultmidpunct}
{\mcitedefaultendpunct}{\mcitedefaultseppunct}\relax
\EndOfBibitem
\bibitem[Houston and Walkinshaw(2013)Houston, and Walkinshaw]{houston2013consensus}
Houston,~D.~R.; Walkinshaw,~M.~D. Consensus docking: improving the reliability of docking in a virtual screening context. \emph{J. Chem. Inform. Model.} \textbf{2013}, \emph{53}, 384--390\relax
\mciteBstWouldAddEndPuncttrue
\mciteSetBstMidEndSepPunct{\mcitedefaultmidpunct}
{\mcitedefaultendpunct}{\mcitedefaultseppunct}\relax
\EndOfBibitem
\bibitem[Jiang \latin{et~al.}(2021)Jiang, Hsieh, Wu, Kang, Wang, Wang, Liao, Shen, Xu, Wu, \latin{et~al.} others]{IGN}
Jiang,~D.; Hsieh,~C.-Y.; Wu,~Z.; Kang,~Y.; Wang,~J.; Wang,~E.; Liao,~B.; Shen,~C.; Xu,~L.; Wu,~J.; others Interactiongraphnet: A novel and efficient deep graph representation learning framework for accurate protein--ligand interaction predictions. \emph{J. Med. Chem.} \textbf{2021}, \emph{64}, 18209--18232\relax
\mciteBstWouldAddEndPuncttrue
\mciteSetBstMidEndSepPunct{\mcitedefaultmidpunct}
{\mcitedefaultendpunct}{\mcitedefaultseppunct}\relax
\EndOfBibitem
\bibitem[Li \latin{et~al.}(2023)Li, Guan, Zhang, Sun, Wang, Bagni, and Head-Gordon]{LP-PDBBind}
Li,~J.; Guan,~X.; Zhang,~O.; Sun,~K.; Wang,~Y.; Bagni,~D.; Head-Gordon,~T. Leak Proof PDBBind: A Reorganized Dataset of Protein-Ligand Complexes for More Generalizable Binding Affinity Prediction. \emph{under review} \textbf{2023}, \relax
\mciteBstWouldAddEndPunctfalse
\mciteSetBstMidEndSepPunct{\mcitedefaultmidpunct}
{}{\mcitedefaultseppunct}\relax
\EndOfBibitem
\bibitem[Brown \latin{et~al.}(2019)Brown, Fiscato, Segler, and Vaucher]{brown2019guacamol}
Brown,~N.; Fiscato,~M.; Segler,~M.~H.; Vaucher,~A.~C. GuacaMol: benchmarking models for de novo molecular design. \emph{J. Chem. Inform. Model.} \textbf{2019}, \emph{59}, 1096--1108\relax
\mciteBstWouldAddEndPuncttrue
\mciteSetBstMidEndSepPunct{\mcitedefaultmidpunct}
{\mcitedefaultendpunct}{\mcitedefaultseppunct}\relax
\EndOfBibitem
\bibitem[Preuer \latin{et~al.}(2018)Preuer, Renz, Unterthiner, Hochreiter, and Klambauer]{preuer2018frechet}
Preuer,~K.; Renz,~P.; Unterthiner,~T.; Hochreiter,~S.; Klambauer,~G. Fr{\'e}chet ChemNet distance: a metric for generative models for molecules in drug discovery. \emph{J. Chem. Inform. Model.} \textbf{2018}, \emph{58}, 1736--1741\relax
\mciteBstWouldAddEndPuncttrue
\mciteSetBstMidEndSepPunct{\mcitedefaultmidpunct}
{\mcitedefaultendpunct}{\mcitedefaultseppunct}\relax
\EndOfBibitem
\bibitem[Boby \latin{et~al.}(2023)Boby, Fearon, Ferla, Filep, Koekemoer, Robinson, Consortium‡, Chodera, Lee, London, \latin{et~al.} others]{boby2023open}
Boby,~M.~L.; Fearon,~D.; Ferla,~M.; Filep,~M.; Koekemoer,~L.; Robinson,~M.~C.; Consortium‡,~C.~M.; Chodera,~J.~D.; Lee,~A.~A.; London,~N.; others Open science discovery of potent noncovalent SARS-CoV-2 main protease inhibitors. \emph{Science} \textbf{2023}, \emph{382}, eabo7201\relax
\mciteBstWouldAddEndPuncttrue
\mciteSetBstMidEndSepPunct{\mcitedefaultmidpunct}
{\mcitedefaultendpunct}{\mcitedefaultseppunct}\relax
\EndOfBibitem
\bibitem[Erlanson \latin{et~al.}(2004)Erlanson, McDowell, and O'Brien]{erlanson2004}
Erlanson,~D.~A.; McDowell,~R.~S.; O'Brien,~T. Fragment-based drug discovery. \emph{J. Med. Chem.} \textbf{2004}, \emph{47}, 3463--3482\relax
\mciteBstWouldAddEndPuncttrue
\mciteSetBstMidEndSepPunct{\mcitedefaultmidpunct}
{\mcitedefaultendpunct}{\mcitedefaultseppunct}\relax
\EndOfBibitem
\bibitem[Beigel \latin{et~al.}(2020)Beigel, Tomashek, Dodd, Mehta, Zingman, Kalil, Hohmann, Chu, Luetkemeyer, Kline, Lopez~de Castilla, Finberg, Dierberg, Tapson, Hsieh, Patterson, Paredes, Sweeney, Short, Touloumi, Lye, Ohmagari, Oh, Ruiz-Palacios, Benfield, Fätkenheuer, Kortepeter, Atmar, Creech, Lundgren, Babiker, Pett, Neaton, Burgess, Bonnett, Green, Makowski, Osinusi, Nayak, and Lane]{Beigel2020}
Beigel,~J.~H. \latin{et~al.}  Remdesivir for the Treatment of Covid-19 - Final Report. \emph{N. Engl. J. Med.} \textbf{2020}, \emph{383}, 1813--1826\relax
\mciteBstWouldAddEndPuncttrue
\mciteSetBstMidEndSepPunct{\mcitedefaultmidpunct}
{\mcitedefaultendpunct}{\mcitedefaultseppunct}\relax
\EndOfBibitem
\bibitem[Unoh \latin{et~al.}(2022)Unoh, Uehara, Nakahara, Nobori, Yamatsu, Yamamoto, Maruyama, Taoda, Kasamatsu, Suto, Kouki, Nakahashi, Kawashima, Sanaki, Toba, Uemura, Mizutare, Ando, Sasaki, Orba, Sawa, Sato, Sato, Kato, and Tachibana]{Unoh2022}
Unoh,~Y. \latin{et~al.}  Discovery of S-217622, a Noncovalent Oral SARS-CoV-2 3CL Protease Inhibitor Clinical Candidate for Treating COVID-19. \emph{J. Med. Chem.} \textbf{2022}, \emph{65}, 6499--6512\relax
\mciteBstWouldAddEndPuncttrue
\mciteSetBstMidEndSepPunct{\mcitedefaultmidpunct}
{\mcitedefaultendpunct}{\mcitedefaultseppunct}\relax
\EndOfBibitem
\bibitem[Li \latin{et~al.}(2021)Li, Michelson, Foraker, Zhan, and Payne]{li2021computational}
Li,~F.; Michelson,~A.~P.; Foraker,~R.; Zhan,~M.; Payne,~P.~R. Computational analysis to repurpose drugs for COVID-19 based on transcriptional response of host cells to SARS-CoV-2. \emph{BMC Med. Inform. Dec. Mak.} \textbf{2021}, \emph{21}, 1--13\relax
\mciteBstWouldAddEndPuncttrue
\mciteSetBstMidEndSepPunct{\mcitedefaultmidpunct}
{\mcitedefaultendpunct}{\mcitedefaultseppunct}\relax
\EndOfBibitem
\bibitem[Hegyi and Ziebuhr(2002)Hegyi, and Ziebuhr]{hegyi2002conservation}
Hegyi,~A.; Ziebuhr,~J. Conservation of substrate specificities among coronavirus main proteases. \emph{J. General Viro.} \textbf{2002}, \emph{83}, 595--599\relax
\mciteBstWouldAddEndPuncttrue
\mciteSetBstMidEndSepPunct{\mcitedefaultmidpunct}
{\mcitedefaultendpunct}{\mcitedefaultseppunct}\relax
\EndOfBibitem
\bibitem[Moghadasi \latin{et~al.}(2023)Moghadasi, Heilmann, Khalil, Nnabuife, Kearns, Ye, Moraes, Costacurta, Esler, Aihara, von Laer, Martinez-Sobrido, Palzkill, Amaro, and Harris]{Moghadasi2023}
Moghadasi,~S.~A.; Heilmann,~E.; Khalil,~A.~M.; Nnabuife,~C.; Kearns,~F.~L.; Ye,~C.; Moraes,~S.~N.; Costacurta,~F.; Esler,~M.~A.; Aihara,~H.; von Laer,~D.; Martinez-Sobrido,~L.; Palzkill,~T.; Amaro,~R.~E.; Harris,~R.~S. Transmissible SARS-CoV-2 variants with resistance to clinical protease inhibitors. \emph{Sci. Adv.} \textbf{2023}, \emph{9}, eade8778\relax
\mciteBstWouldAddEndPuncttrue
\mciteSetBstMidEndSepPunct{\mcitedefaultmidpunct}
{\mcitedefaultendpunct}{\mcitedefaultseppunct}\relax
\EndOfBibitem
\bibitem[Ventola(2015)]{Ventola2015}
Ventola,~C.~L. {The antibiotic resistance crisis: part 1: causes and threats.} \emph{P \& T : a peer-reviewed journal for formulary management} \textbf{2015}, \emph{40}, 277--83\relax
\mciteBstWouldAddEndPuncttrue
\mciteSetBstMidEndSepPunct{\mcitedefaultmidpunct}
{\mcitedefaultendpunct}{\mcitedefaultseppunct}\relax
\EndOfBibitem
\bibitem[Polyak and Juditsky(1992)Polyak, and Juditsky]{polyak1992acceleration}
Polyak,~B.~T.; Juditsky,~A.~B. Acceleration of stochastic approximation by averaging. \emph{SIAM J. Contr. Opt.} \textbf{1992}, \emph{30}, 838--855\relax
\mciteBstWouldAddEndPuncttrue
\mciteSetBstMidEndSepPunct{\mcitedefaultmidpunct}
{\mcitedefaultendpunct}{\mcitedefaultseppunct}\relax
\EndOfBibitem
\bibitem[Wan \latin{et~al.}(2013)Wan, Zeiler, Zhang, Le~Cun, and Fergus]{wan2013regularization}
Wan,~L.; Zeiler,~M.; Zhang,~S.; Le~Cun,~Y.; Fergus,~R. Regularization of neural networks using dropconnect. International conference on machine learning. 2013; pp 1058--1066\relax
\mciteBstWouldAddEndPuncttrue
\mciteSetBstMidEndSepPunct{\mcitedefaultmidpunct}
{\mcitedefaultendpunct}{\mcitedefaultseppunct}\relax
\EndOfBibitem
\bibitem[Paszke \latin{et~al.}(2017)Paszke, Gross, Chintala, Chanan, Yang, DeVito, Lin, Desmaison, Antiga, and Lerer]{paszke2017automatic}
Paszke,~A.; Gross,~S.; Chintala,~S.; Chanan,~G.; Yang,~E.; DeVito,~Z.; Lin,~Z.; Desmaison,~A.; Antiga,~L.; Lerer,~A. Automatic differentiation in PyTorch. \textbf{2017}, \relax
\mciteBstWouldAddEndPunctfalse
\mciteSetBstMidEndSepPunct{\mcitedefaultmidpunct}
{}{\mcitedefaultseppunct}\relax
\EndOfBibitem
\bibitem[Howard and Gugger(2020)Howard, and Gugger]{howard2020fastai}
Howard,~J.; Gugger,~S. Fastai: a layered API for deep learning. \emph{Information} \textbf{2020}, \emph{11}, 108\relax
\mciteBstWouldAddEndPuncttrue
\mciteSetBstMidEndSepPunct{\mcitedefaultmidpunct}
{\mcitedefaultendpunct}{\mcitedefaultseppunct}\relax
\EndOfBibitem
\bibitem[Kingma and Ba(2014)Kingma, and Ba]{kingma2014adam}
Kingma,~D.~P.; Ba,~J. Adam: A method for stochastic optimization. \emph{arXiv preprint arXiv:1412.6980} \textbf{2014}, \relax
\mciteBstWouldAddEndPunctfalse
\mciteSetBstMidEndSepPunct{\mcitedefaultmidpunct}
{}{\mcitedefaultseppunct}\relax
\EndOfBibitem
\bibitem[Smith(2018)]{smith2018disciplined}
Smith,~L.~N. A disciplined approach to neural network hyper-parameters: Part 1--learning rate, batch size, momentum, and weight decay. \emph{arXiv preprint arXiv:1803.09820} \textbf{2018}, \relax
\mciteBstWouldAddEndPunctfalse
\mciteSetBstMidEndSepPunct{\mcitedefaultmidpunct}
{}{\mcitedefaultseppunct}\relax
\EndOfBibitem
\bibitem[Smith and Topin(2019)Smith, and Topin]{smith2019super}
Smith,~L.~N.; Topin,~N. Super-convergence: Very fast training of neural networks using large learning rates. Artificial Intelligence and Machine Learning for Multi-Domain Operations Applications. 2019; p 1100612\relax
\mciteBstWouldAddEndPuncttrue
\mciteSetBstMidEndSepPunct{\mcitedefaultmidpunct}
{\mcitedefaultendpunct}{\mcitedefaultseppunct}\relax
\EndOfBibitem
\bibitem[Williams(1992)]{williams1992simple}
Williams,~R.~J. Simple statistical gradient-following algorithms for connectionist reinforcement learning. \emph{Machine learning} \textbf{1992}, \emph{8}, 229--256\relax
\mciteBstWouldAddEndPuncttrue
\mciteSetBstMidEndSepPunct{\mcitedefaultmidpunct}
{\mcitedefaultendpunct}{\mcitedefaultseppunct}\relax
\EndOfBibitem
\end{mcitethebibliography}

\begin{tocentry}
\includegraphics{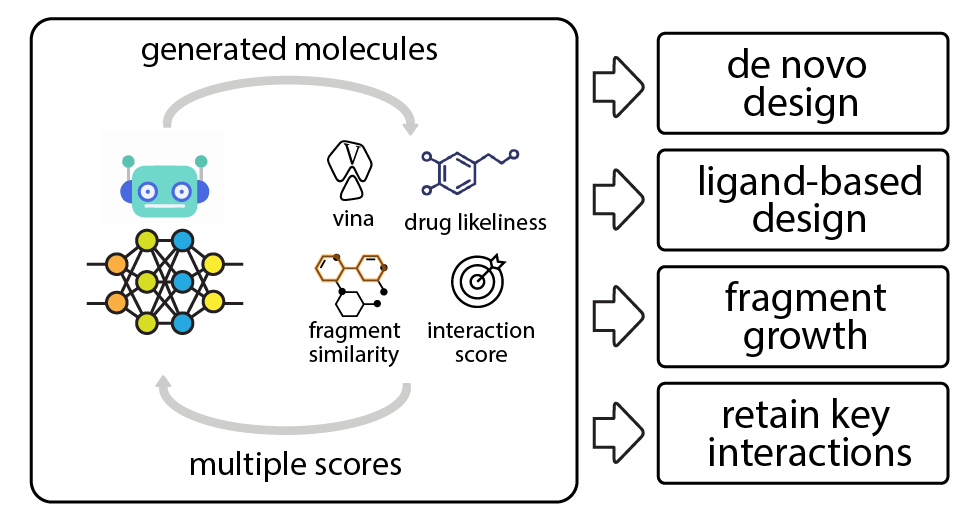}
\label{figure:toc}
\end{tocentry}
\end{document}


\section{\fontsize{16}{16}\selectfont Methodology Details}

\textbf{Tokens in the generative model}. Here we provide a complete list of tokens used in the generative model:
\begin{itemize}
    \item SELFIES tokens: `C', `=C', `Ring1', `Branch1', `N', `=Branch1', `O', `=O', `Ring2', `Branch2', `=N', `S', `\#Branch1', `=Branch2', `F', `\#Branch2', `C@H1', `C@@H1', `\#C', `Cl', `P', `/C', `NH1', `=Ring1', `C@', `C@@', `O-1', `Br', `N+1', `\#N', `\verb|\\|C', `=Ring2', `/N', `=S', `=N+1', `\verb|\\|N', `I', `/O', `\verb|\\|O', `\verb|\\|S', `S+1', `/S', `/C@@H1', `/C@H1', `Se', `=N-1', `=P', `N-1', `\verb|\\|C@@H1', `\verb|\\|C@H1', `/N+1', `C-1', `\#N+1', `P+1', `\verb|\\|NH1', `OH0', `/Br', `P@', `P@@', `\verb|\\|Cl', `\verb|\\|O-1', `\verb|\\|N+1', `/C@@', `-/Ring2', `/Cl', `/C@', `PH1', `=Se'
    \item Functional tokens: ``Break'
\end{itemize}
When sampling molecules represented as SELFIES strings, the first token was always selected as the ``Break' token. Then each token was sampled with probability distribution predicted by the generative model. Once the ``Break' token was selected again, or the total number of tokens exceeded 500, a single molecule sampling process was considered complete. Sampled strings containing invalid atom types (Si, B \& Te) in Autodock Vina\cite{Trott2010} were discarded as well.
\newline

\noindent
\textbf{Supervised fine-tuning of the network for specialized tasks}. The same weight decay and dropout multiplier as in the pre-training with ChEMBL 24\cite{Gaulton2012} were used in this stage. The pre-trained model was additionally tuned for 5-20 epochs to condition the molecule generation on the target substructures yet with structural diversity. For the task to generate analogs to the inhibitors reported by Zhang \textit{et al.}\cite{zhang2021potent}, the model was fine-tuned with 14 SMILES strings from PDB deposits (PDB ID 7L10, 7L11, 7L12, 7L13, 7L14, 7M8M, 7M8N, 7M8O, 7M8P, 7M8X, 7M8Y, 7M8Z, 7M90, 7M91). For the fragment growth design task, the fine-tuning set consisted of 867 molecules with substructure match from SciFinder, excluding the ones reported by Zhang \textit{et al.}.
\newline

\noindent
\textbf{Mpro Active Site Receptor Preparation and Docking Procedures.}
Stzain et al.\cite{Sztain2020} simulated SARS-CoV-2 Mpro (PDB ID 6LU7\cite{Jin2020}) with Gaussian Accelerated Molecular Dynamics to characterize active site and dimer interface dynamics, as well as elucidate the presence of cryptic binding pockets. In total, Sztain et al. produced 6 microseconds of enhanced-sampled Mpro conformations.\cite{Sztain2020} These extensive simulations represent an invaluable resource for SARS-CoV-2 antiviral design, and as such Sztain et al. shared their trajectories publicly (\url{https://amarolab.ucsd.edu/covid19.php}) in accordance with the data sharing philosophy put forth by Amaro and Mulholland.\cite{Amaro2020} To ensure we were selecting biologically relevant Mpro conformations for use in our molecule generation workflow, we selected receptor structures from Sztain et al.'s simulations for the \textit{de novo} task. For compatibility of the generated analogs and fragment-derived molecules, we used SARS-CoV-2 conformation from the protein-ligand co-crystal structure (PDB ID 7L11\cite{zhang2021potent}) as the receptor for docking in the relevant tasks. Selection of each receptor structure and subsequent protein preparation steps are described below. 

To generate molecules targeting the Mpro active site, we selected the representative structure from the most populated cluster identified in Sztain \textit{et al.}'s enhanced sampling trajectories of Mpro dimer,\cite{Sztain2020} simulated with a covalently bound inhibitor called N3. From Sztain et al.'s freely available files, the filename of the selected protease structure was ``\url{5.0_2.0_147.0_147.0_295.0_c0.pdb}''. We deleted the covalently bound N3 from this structure, taking care not to delete the catalytic Cysteine atoms (resids 145 and 451). We then modified the C145 and C451 atom names so that they reflected canonical Cysteine atom names. The cartesian coordinates for the active site center were found by calculating the center of mass of the C145 bound N3I covalent inhibitor before the inhibitor was deleted ([atomselect top ``resname N3I and resid 145'']). This center of mass (x=54.58, y=45.92, z=75.06) was used to define the center of the active site during receptor grid generation steps in AutoDock Vina. We aligned the active site of Zhang \textit{et al.}'s co crystal structure to the Sztain's such that the same receptor grid center can be applied. 

AutoDockFR\cite{ravindranath2015autodockfr} was used to convert Mpro \url{.pdb} files to AutoDock Vina\cite{Trott2010} compatible \url{.pdbqt} files. Additionally, Meeko and Open Babel\cite{openbabel} were used to convert generated molecule structure files to AutoDock Vina compatible \url{.pdbqt} files. Gastieger charges were used for all AutoDock Vina structures. A cubic receptor grid of 24\r{A} x 24\r{A} x 24\r{A} was centered around binding site's central coordinate (listed above), with a grid spacing of 1.0\r{A}.
\newline



\newpage
\section{\fontsize{16}{16}\selectfont Supporting Tables}

\begin{longtable}[c]{lp{0.5\textwidth}cc}
    \caption{The filtered sets of molecules using iMiner structure based analogs generation}

 \endfirsthead

 \multicolumn{4}{l}{Table S1 (continued)}\\
 \hline
 \hline
 Index&Canonical SMILES& Vina score & IGN score \\
 \hline
 \endhead

 \hline
 \endfoot

 \endlastfoot
\hline\hline
            Index&Canonical SMILES& newVina score & newIGN score \\ \hline
\multicolumn{4}{l}{Structural Analogs} \\
1&\mbox{CCC=C(c1cc(Cl)cc(-c2cc(C3CNC(=O)NC3=O)cn(-c3ccccn3)c2=O)c1)c1ccc(O)nn1} \\
&&-11.81 & -10.68  \\
2&\mbox{CC(C)CC(N)C1CN(C(=O)Nc2cc(Cl)cc(-c3cc(-c4c[nH]c(=O)[nH]c4=O)cn} \\
&(-c4cccnc4)c3=O)c2)CCN1&-11.82 & -10.97 \\
3&\mbox{Cc1ccccc1CNC(=O)C(C)(C)Oc1cc(Cl)cc(-c2cc(-c3n[nH]c(=S)[nH]c3=O)cn} \\
&(-c3cccnc3)c2=O)c1&-11.46 & -10.76   \\
4&\mbox{O=C1NC(=Cc2n[nH]cc2Br)NC=C1c1cc(-c2cc(Cl)cc(OCc3cccc(F)c3F)c2)c(=O)n} \\
&(-c2cccnc2)c1&-10.94 & -10.82   \\
5&\mbox{CNCC1CN(CCOc2cc(Cl)cc(-c3cc(-c4c[nH]c(=N)[nH]c4=O)cn(-c4cccnc4)c3=O)c2)CCN1} \\
&&-11.82 & -10.79  \\
6&\mbox{O=c1[nH]cc(-c2cc(-c3cc(Cl)cc(CC=Cc4ccccc4)c3)c(=O)n(-c3cccnc3)c2)c(=O)[nH]1} \\
&&-10.90 & -10.75  \\
7&\mbox{CS(=O)(=O)NC1CN(CCOc2cc(Cl)cc(-c3cc(-c4c[nH]c(=O)[nH]c4=O)cn} \\
&(-c4cccnc4)c3=O)c2)CCN1&-11.73 & -10.87  \\
8&\mbox{O=C1C=CC=C(COc2cc(Cl)cc(-c3cc(-c4c[nH]c(=O)[nH]c4=O)cn(-c4cccnc4)c3=O)c2)C1} \\
&&-11.68 & -10.71\\
9&\mbox{O=c1[nH]cc(-c2cc(-c3cc(Cl)cc(OCC4=CC=CC(F)C4)c3)c(=O)n(-c3cccnc3)c2)c(=O)[nH]1} \\
&&-11.47 & -10.70  \\
10&\mbox{O=c1[nH]cc(-c2cc(-c3cc(Cl)cc(OCc4cccc(F)c4)n3)c(=O)n(-c3cccnc3)c2)c(=O)[nH]1} \\
&&-11.14 & -10.75 \\
11&\mbox{O=c1[nH]cc(-c2cc(-c3cc(Cl)c(F)cc3OCc3cccc(Cl)c3)c(=O)n(-c3cccnc3)c2)c(=O)[nH]1} \\
&&-11.28 & -10.85  \\
12&\mbox{O=c1[nH]cc(-c2cc(-c3cc(Cl)cc(OCc4cccc(F)c4F)c3)c(=O)n(-c3ccccn3)c2)c(=O)[nH]1} \\
&&-11.60 & -10.74  \\
13&\mbox{O=c1[nH]cc(-c2cc(-c3cc(F)c(F)c(OCc4ccccc4Cl)c3)c(=O)n(-c3cccnc3)c2)c(=O)[nH]1} \\
&&-11.68 & -10.76   \\
14&\mbox{O=c1[nH]cc(-c2cc(-c3cc(Cl)cc(OCc4cccc(F)c4)c3)c(=O)n(-c3ccccn3)c2)c(=O)[nH]1} \\
&&-12.32 & -10.68  \\
15&\mbox{N\#Cc1ccccc1Oc1cc(F)cc(-c2cc(-c3c[nH]c(=O)[nH]c3=O)cn(-c3cccnc3)c2=O)c1} \\
&&-11.72 & -10.68  \\
16&\mbox{O=c1cc(COc2cc(Cl)cc(-c3cc(-c4c[nH]c(=O)[nH]c4=O)cn(-c4cccnc4)c3=O)c2)cn[nH]1} \\
&&-10.95 & -10.72  \\
17&\mbox{O=Cc1c(-c2c[nH]c(=O)[nH]2)cc(-c2cc(Br)c(F)c(OCc3ccccc3Cl)c2)c(=O)n1-c1cccnc1} \\
&&-11.31 & -10.74   \\
18&\mbox{O=c1[nH]cc(-c2cc(-c3cc(Cl)cc(OCc4ccccc4Cl)c3)c(=O)n(-c3cccnc3)c2)c(=S)[nH]1} \\
&&-11.10 & -10.74   \\
19&\mbox{O=c1[nH]c(=S)[nH]cc1-c1cc(-c2cc(Cl)cc(OCc3cccc(F)c3)c2)c(=O)n(-c2cccnc2)c1} \\
&&-11.14 & -10.68 \\
20&\mbox{N=c1[nH]cc(-c2cc(-c3cc(Cl)cc(OCc4cccc(F)c4)c3)c(=O)n(-c3cccnc3)c2)c(=O)[nH]1} \\
&&-12.01 & -10.71  \\
21&\mbox{O=c1[nH]cc(-c2cc(-c3cc(Cl)cc(OCc4ccnc(F)c4)c3)c(=O)n(-c3cccnc3)c2)c(=O)[nH]1} \\
&&-11.48 & -10.75  \\
22&\mbox{O=c1[nH]cc(-c2cc(-c3cc(Cl)cc(OCc4ccc(F)nc4)c3)c(=O)n(-c3cccnc3)c2)c(=O)[nH]1} \\
&&-11.15 & -10.74  \\
23&\mbox{O=c1[nH]cc(-c2cc(-c3cc(Cl)cc(OCc4cncc(F)c4)c3)c(=O)n(-c3cccnc3)c2)c(=O)[nH]1} \\
&&-11.31 & -10.72 \\
24&\mbox{O=c1[nH]cc(-c2cc(-c3cc(Cl)c(F)c(OCc4ccccc4F)c3)c(=O)n(-c3cccnc3)c2)c(=O)[nH]1} \\
&&-11.91 & -10.81  \\
25&\mbox{Nc1cc(-c2cc(-c3c[nH]c(=O)[nH]c3=O)cn(-c3cccnc3)c2=O)cc(OCc2ccccc2Cl)c1F} \\
&&-11.71 & -10.75  \\
26&\mbox{O=c1[nH]cc(-c2cc(-c3cc(I)cc(OCc4cccc(F)c4)c3)c(=O)n(-c3cccnc3)c2)c(=O)[nH]1} \\
&&-10.89 & -10.91   \\
        \hline
    \hline
\label{tab:guacamol1}
\end{longtable}

\begin{longtable}[c]{lp{0.5\textwidth}cc}
    \caption{The filtered sets of molecules using iMiner structure based fragment growth}

 \endfirsthead

 \multicolumn{4}{l}{Table S2 (continued)}\\
 \hline
 \hline
 Index&Canonical SMILES& Vina score & IGN score \\
 \hline
 \endhead

 \hline
 \endfoot

 \endlastfoot
\hline\hline
            Index&Canonical SMILES& newVina score & newIGN score \\ \hline
\multicolumn{4}{l}{Structure-based Fragment Growth} \\
1&\mbox{COc1ccc2cc(-c3ccc4nn(C)cc4c3)c(=O)n(-c3ccc(Cl)nc3)c2n1} \\
&&-9.48 & -10.48   \\
2&\mbox{O=c1c(-c2ccc3ncccc3c2)cc(-c2ccccn2)cn1-c1ccc(F)nc1} \\
&&-10.07 & -10.49   \\
3&\mbox{COc1ncc(-n2c(=O)c(-c3ccc4nn(C)cc4c3)cc3ccc(NCC(F)(F)F)nc32)cc1N} \\
&&-10.47 & -10.56   \\
4&\mbox{CNc1ccc2cc(-c3ccc4nn(C)cc4c3)c(=O)n(-c3ccc(N)nc3)c2n1} \\
&&-10.11 & -10.43   \\
5&\mbox{C[C@H](NC(=O)c1cc(-c2ccc(F)cc2)c(=O)n(-c2cccnc2)c1)c1cnc(C(F)(F)F)nc1} \\
&&-9.55 & -10.60   \\
6&\mbox{COc1ccc(-n2c(=O)c3c[nH]nc3c3cnc(NC[C@@H]4CN(C)c5ccccc5O4)nc32)cn1} \\
&&-9.66 & -10.64   \\
7&\mbox{C[C@H](NC(=O)c1c(N)nn2cccnc12)c1cc2cccc(Cl)c2c(=O)n1-c1cncc(F)c1} \\
&&-9.75 & -10.73   \\
8&\mbox{CC(C)C(=O)N[C@H]1C=Cc2nc3c4ccccc4c(=O)n(-c4ccc(Cl)nc4)c3n2O1} \\
&&-9.64 & -10.53  \\
9&\mbox{CNc1ccc(-n2c(=O)c(-c3ccc4nn(C)nc4c3)cc3ccc(OCC(F)(F)F)nc32)cn1} \\
&&-10.48 & -10.59   \\
10&\mbox{Cn1cc2cc(-c3cc4ccc(OCC(F)(F)F)nc4n(-c4ccc(N)nc4)c3=O)cnc2n1} \\
&&-10.05 & -10.52  \\
11&\mbox{CNc1ccc(-n2c(=O)c(C(=O)NCC3=CC(C)=NC3)cc3ccc(OCC(F)(F)F)nc32)cn1} \\
&&-9.57 & -10.45   \\
12&\mbox{O=C(O)CCC(=O)Nc1ccc2nc3c4ccccc4c(=O)n(-c4ccc(Cl)nc4)c3n2n1} \\
&&-10.04 & -10.51  \\
        \hline
    \hline
\label{tab:guacamol2}
\end{longtable}

\begin{longtable}[c]{lp{0.5\textwidth}cc}
    \caption{The filtered sets of molecules using iMiner ligand-protein interaction generation}

 \endfirsthead

 \multicolumn{4}{l}{Table S3 (continued)}\\
 \hline
 \hline
 Index&Canonical SMILES& Vina score & IGN score \\
 \hline
 \endhead

 \hline
 \endfoot

 \endlastfoot
\hline\hline
            Index&Canonical SMILES& newVina score & newIGN score \\ \hline
\multicolumn{4}{l}{Interaction-based} \\
1&\mbox{Nc1ncc(-c2cccc(-c3nnn[nH]3)c2)cc1C=C1C=CC(c2cccc3ccccc23)=C1Cl} \\
&&-9.66 & -10.26 \\
2&\mbox{Nc1ncc(-c2cccc(-c3nnn[nH]3)c2)cc1C1=CC=CC2[CH]C=CC=C12} \\
&&-9.70 & -10.00\\
3&\mbox{Nc1ncc(-c2cccc(-c3nnn[nH]3)c2)cc1C=c1cccc2c1=CC=CC2} \\
&&-9.61 & -10.20\\
4&\mbox{Nc1ncc(-c2cccc(-c3nnn[nH]3)c2)cc1C=CC1=CC[C]c2ccccc21} \\
&&-9.88 & -10.11\\
5&\mbox{Nc1ncc(C2=CCNC(c3nnn[nH]3)=C2)cc1-c1cccc2ccccc12} \\
&&-9.39 & -9.79\\
6&\mbox{Nc1ncc(-c2cccc(-c3nnn[nH]3)c2)cc1-c1ccnc2ccccc12} \\
&&-10.02 & -10.14\\
7&\mbox{Nc1ncc(-c2cccc(-c3nnn[nH]3)c2)cc1-c1cccc2ccccc12} \\
&&-9.45 & -9.97\\
8&\mbox{COC=C1C=CC(C(=O)Nc2cccc(-c3cccc(-c4nc5ccccc5[nH]4)c3)c2)=C1} \\
&&-9.58 & -10.05 \\
9&\mbox{CNc1nc(-c2ccccc2NC(=O)c2ccc3[nH]c(C)nc3c2)c2ccccc2n1} \\
&&-9.57 & -10.14 \\
10&\mbox{N=C(N)C=C1C=CC(c2nc3ccccc3c(=O)n2-c2ccc3ccccc3c2)=C1} \\
&&-9.91 & -10.11 \\
11&\mbox{Nc1nc(-c2ccc(-c3cccc4ccccc34)o2)c(-c2ccc(F)cc2)c(=O)[nH]1} \\
&&-9.94 & -10.02 \\
    \hline
\label{tab:guacamol}
\end{longtable}

\begin{table}[h]
    \centering
    \caption{RL training details. The hyperparameters include batch size (bs), learning rate (lr), $\alpha$ and $\epsilon$ in the PPO loss function, and reward weights for Vina score $\lambda_{vina}$, drug-likeliness $\lambda_{DL}$, fragment similarity $\lambda_{frag}$, pharmacophore similarity $\lambda_{phm}$ and interaction $\lambda_{interact}$.}
    \begin{tabular}{lccccccccc}
        \hline\hline
        \textbf{Task} & bs & lr & $\alpha$ & $\epsilon$ & $\lambda_{vina}$ & $\lambda_{DL}$ & $\lambda_{frag}$ & $\lambda_{phm}$ & $\lambda_{interact}$ \\ 
        \hline
        \textbf{Unconditional} & 1024 & 0.0001 & 0.01 & 0.1 & 4 & 1 & & &\\
        \textbf{\textit{de novo} generation} & & & & & & & & & \\
        \textbf{Structural analog} & 1024 & 0.0001 & 0.015 & 0.1 & 2 & 1 & 4 & 2 & \\
        \textbf{generation} & & & & & & & & & \\
        \textbf{Structure-based} & 1024 & 0.0001 & 0.012 & 0.1 & 2 & 1 & 4 & & \\
        \textbf{fragment growth} & & & & & & & & & \\
        \textbf{Interaction-based} & 1024 & 0.0001 & 0.012 & 0.1 & 2 & 1 & & & 5 \\
        \textbf{generation} & & & & & & & & & \\
        \hline
    \end{tabular}
\end{table}

\newpage
\section{\fontsize{16}{16}\selectfont Supporting Figures}
\begin{figure*}
\centering
\includegraphics[width=0.95\textwidth]{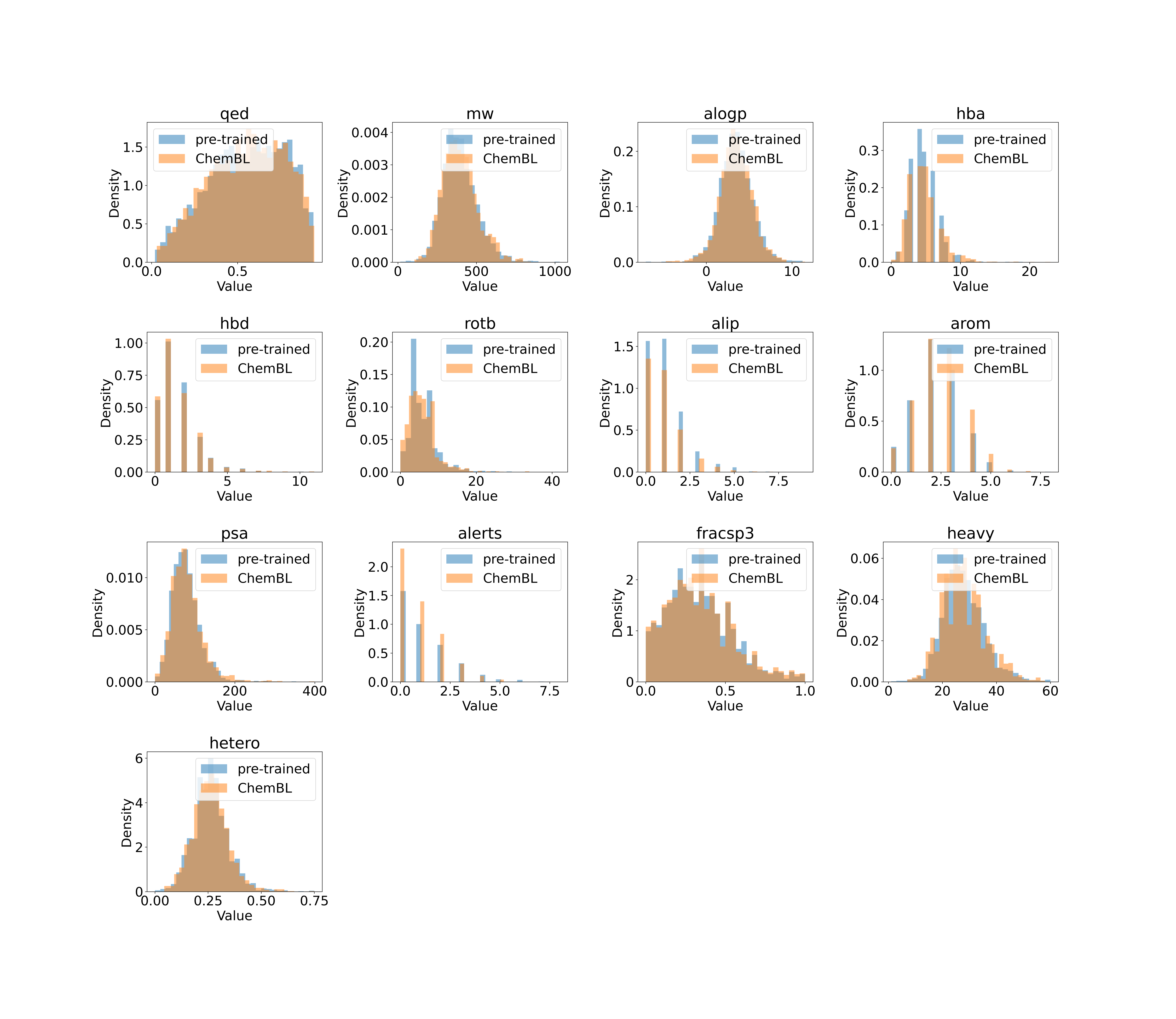}
\vspace{-15mm}
\caption{\textit{Distribution comparisons for 13 different properties of the generated molecules from the pretrained model with molecules from the training dataset (ChEMBL).} The molecular properties considered are well-recognized chemical features related to the drug-likeliness of a molecule which can be obtained through 2D topological connectivity of the molecule: fraction of $sp^3$ hybridized carbons(fracsp3), number of heavy atoms(heavy), fraction of non-carbon atoms in all heavy atoms(hetero), number of hydrogen bond donors(hbd) and acceptors(hba), number of rotatable bonds(rotb), number of aliphatic(alip) and aromatic rings(arom), molecular weight(mw), quantitative estimate of drug-likelihood (QED) value\cite{bickerton2012quantifying}, approximate log partition coefficient between octanol and water (alogP)\cite{wildman1999prediction}, polarizable surface area (PSA), and the number of structural alerts(alerts).\cite{brenk2008lessons} }
\label{fig:prior_comparison}
\end{figure*}

\vspace{-15mm}

\begin{figure}
\centering
\includegraphics[width=0.99\textwidth]{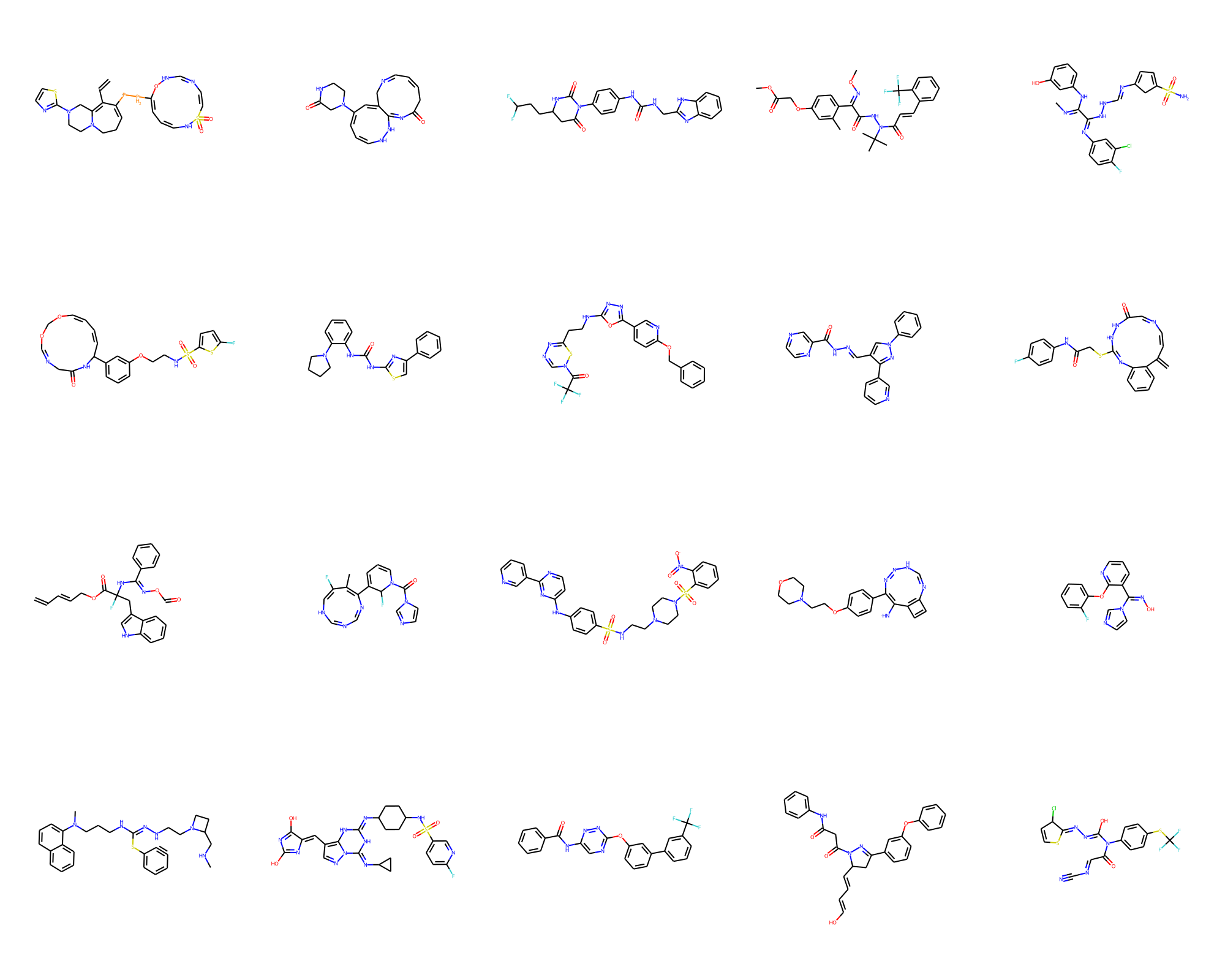}
\caption{Example molecules generated using reinforcement learning without utilizing the drug-likeliness metric as additional reward. Many of these molecules are not drug-like, i.e. having large rings, or having a high proportion of hetero atoms.}
\label{fig:molecules_without_DL}
\end{figure}

\begin{figure*}
    \centering
    \includegraphics[width=0.99\textwidth]{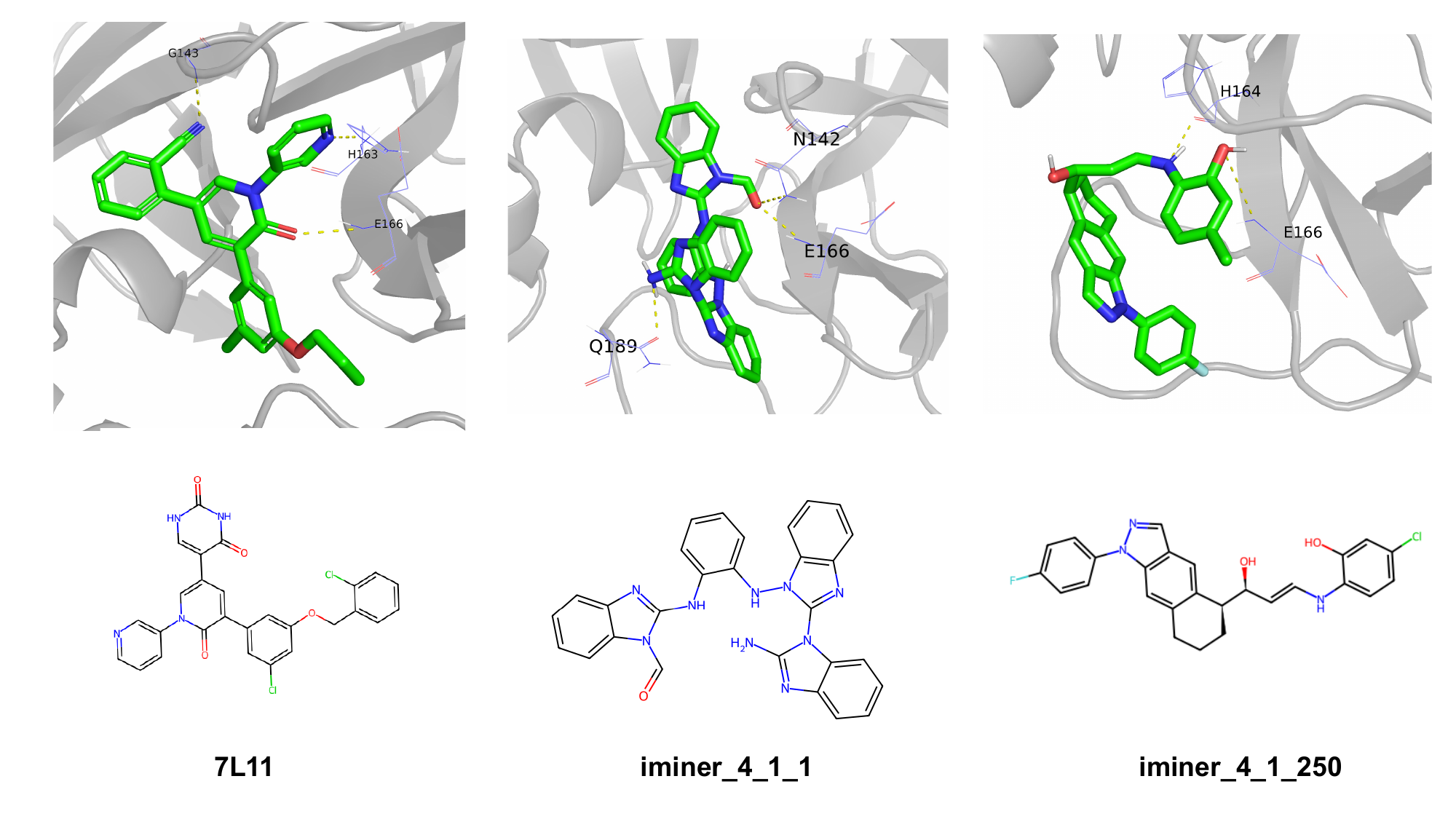}
    \includegraphics[width=0.99\textwidth]{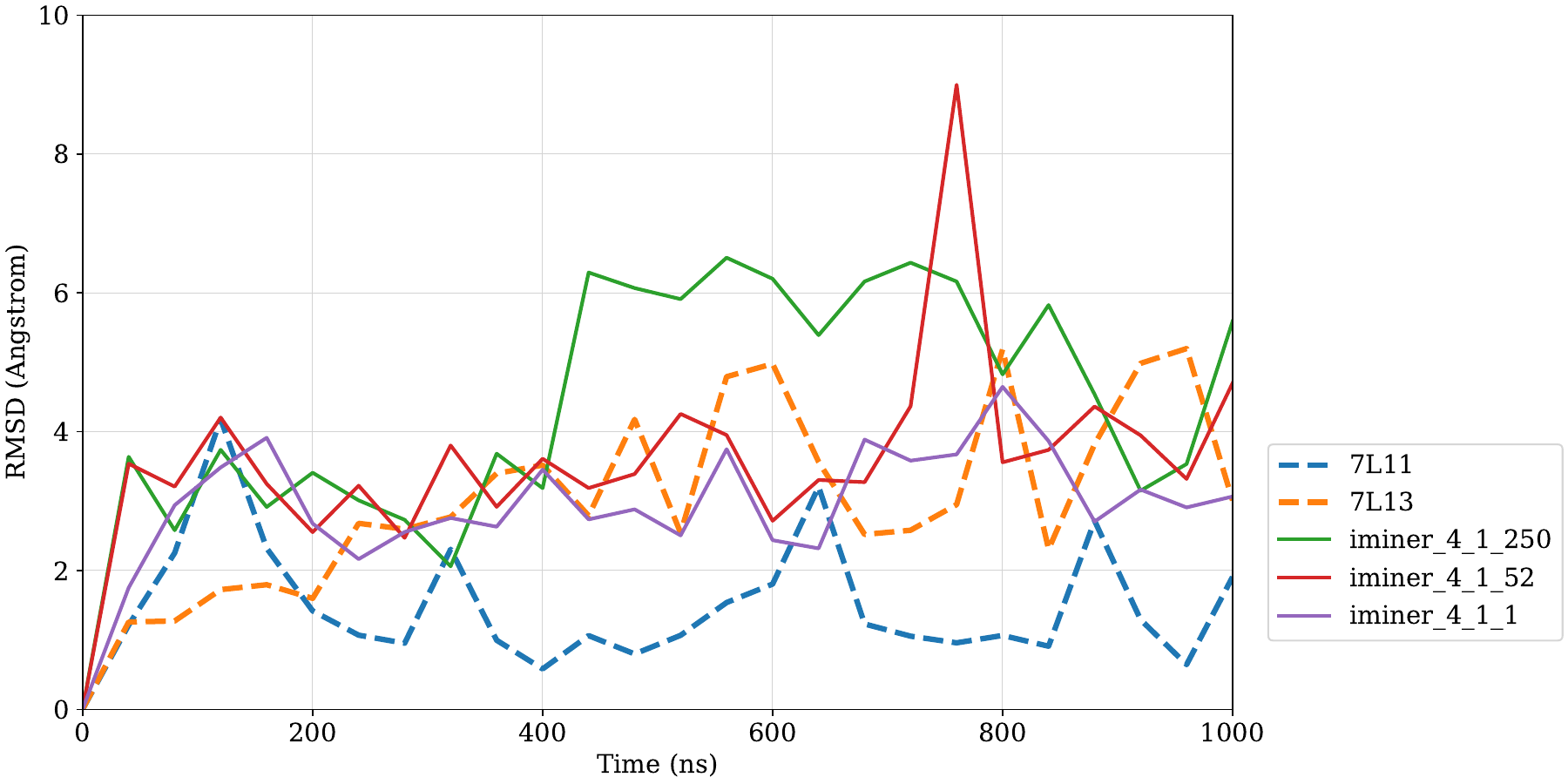}
    \caption{\textit{iMiner workflow for post-analysis.} (a) The binding pose of 7L11 and iMiner 4-1-1 and iMiner 4-1-250. (b) The RMSD of 3 molecules from the unconditional \textit{de novo} design and 2 known binders (PDB code: 7L11 and 7L13) over 1 $\mu$s simulation. (b) }
    \label{fig:task1_md}
\end{figure*}

\begin{figure*}
    \centering
    \includegraphics[width=0.99\textwidth]{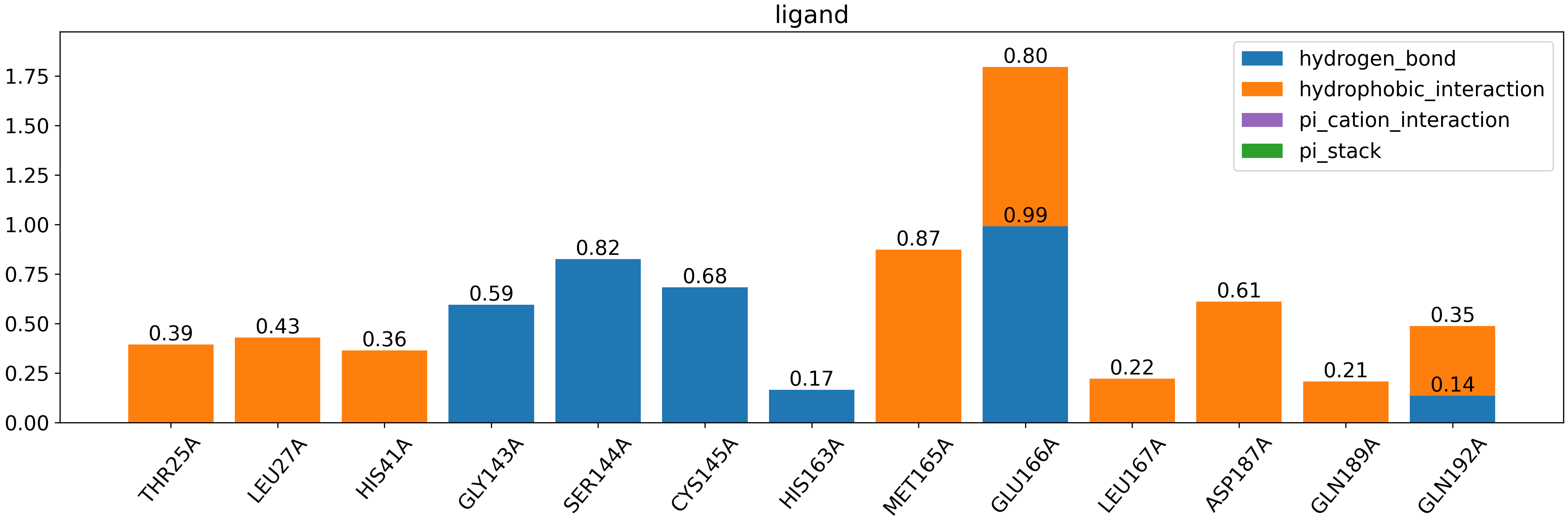}
    \includegraphics[width=0.99\textwidth]{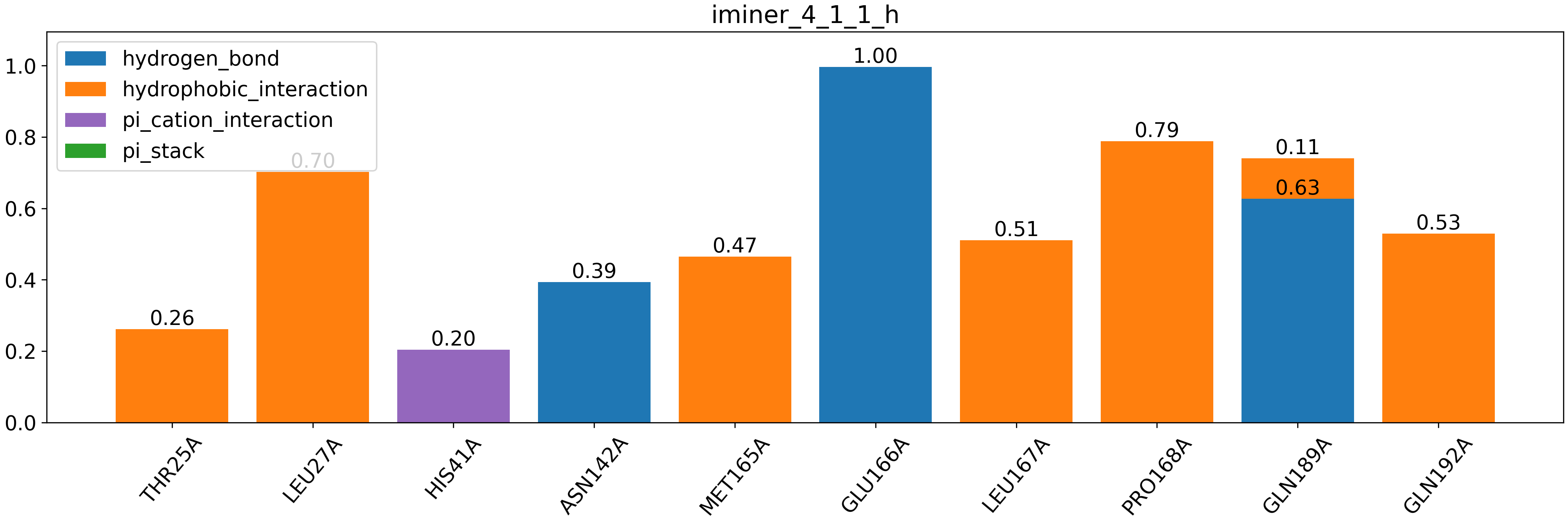}
    \includegraphics[width=0.99\textwidth]{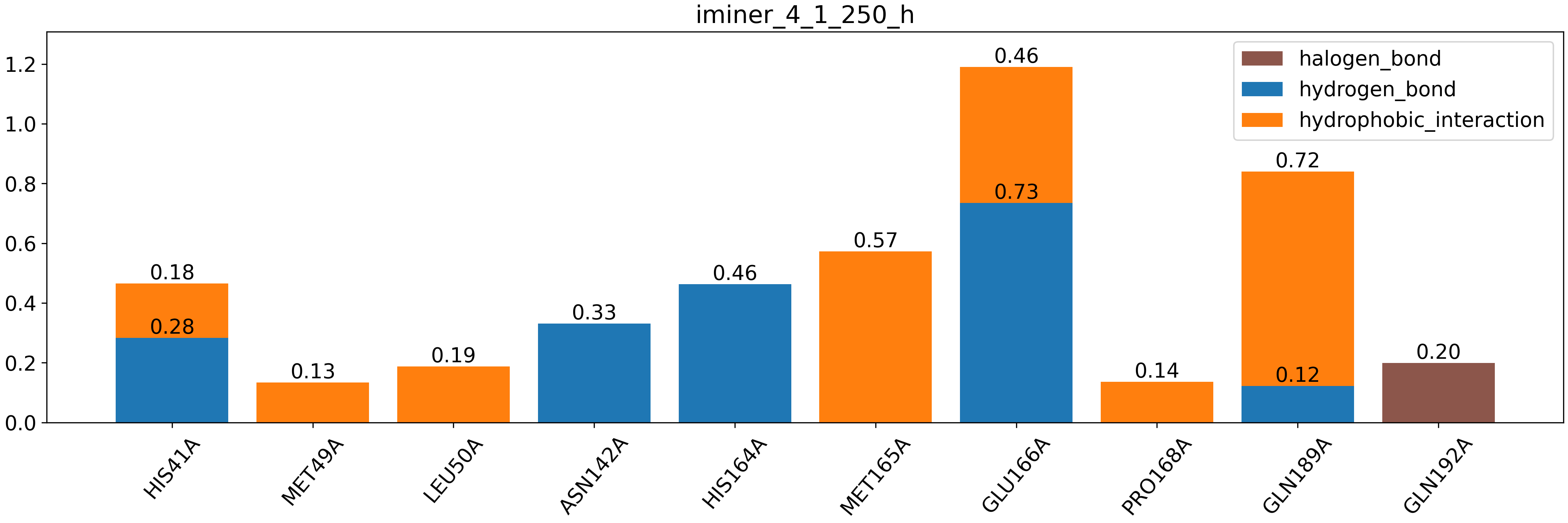}
    \caption{\textit{The molecular interactions between ligand and protein and their frequency over 1$\mu$s MD simulation.} (top) PDB code 7L11; (middle) iMiner 4-1-1; (bottom) iMiner 4-1-250.}
    \label{fig:task1_md2}
\end{figure*}
\newpage

\bibliographystyle{chem-acs}
\bibliography{references}